\documentclass[aps,prb,longbibliography,reprint,groupedaddress,showpacs,amsmath,amssymb,superscriptaddress]{revtex4-2}
\usepackage{graphicx}
\usepackage{caption}
\usepackage{dcolumn}
\usepackage{bm}
\usepackage{float}
\usepackage{url}
\usepackage{miller}
\usepackage{subfigure}
\usepackage{calc}
\usepackage{latexsym}
\usepackage{epsf}
\usepackage{epsfig}
\usepackage{notoccite}
\usepackage{euscript}
\usepackage{hyperref}
\usepackage{xcolor}
\usepackage{sidecap}
\usepackage{gensymb}
\usepackage[english]{babel}
\usepackage[export]{adjustbox}
\usepackage[autostyle]{csquotes} 

\begin{document}
\title{Large Magnetic Anisotropy and Magnetostriction in Thin Films of CoV$_2$O$_4$  }

 \author{Sangsoo Kim} \altaffiliation{These authors contributed equally to this work}
 \affiliation{National High Magnetic Field Laboratory, Florida State University, Tallahassee, FL 32310, USA}
  \affiliation{Department of Physics, Florida State University, Tallahassee, FL 32306, USA}
 \author{Christie Thompson}\altaffiliation{These authors contributed equally to this work}
 \affiliation{National High Magnetic Field Laboratory, Florida State University, Tallahassee, FL 32310, USA}
 \affiliation{Materials Science and Engineering Program, Florida State University, Tallahassee, FL 32310, USA}
 \author{Yan Xin}
  \affiliation{National High Magnetic Field Laboratory, Florida State University, Tallahassee, FL 32310, USA}
 \author{Christianne Beekman}\email{Corresponding author: beekman@magnet.fsu.edu}
 \affiliation{National High Magnetic Field Laboratory, Florida State University, Tallahassee, FL 32310, USA}
  \affiliation{Department of Physics, Florida State University, Tallahassee, FL 32306, USA}
\date{\today}

\begin{abstract}
\noindent 
Spinel Cobalt Vanadate CoV$_2$O$_4$ has been grown on \hkl(001) SrTiO$_3$ substrates. Using torque magnetometry experiments, we find that the previously observed temperature induced anisotropy change, where the easy axis changes from the out of plane \hkl[001] direction to a biaxial anisotropy with planar \hkl<100> easy axes, occurs in a gradual second-order structural phase transition. This work characterizes this transition and the magnetic anisotropies in the \hkl(001), \hkl(100), and \hkl(-110) rotation planes, and explores their field dependence up to 30~T. Below 80~K, hysteretic features appear around the hard axes, i.e., the out-of-plane direction in \hkl(-110) and \hkl(010) rotations and the planar \hkl<110> directions in  \hkl(001) rotations. This is due to a Zeeman Energy that originates from the lag of the magnetization with respect to the applied magnetic field as the sample is rotated. The appearance of the hysteresis, which persist up to very high fields, shows that the anisotropy at low temperature is rather strong. Additionally, field dependent distortions to the symmetry of the torque response in increasing applied fields shows that magnetostriction plays a large role in determining the direction and magnitude of the anisotropy.
\end{abstract}
\maketitle
\section{INTRODUCTION}
The strongly correlated electron community has extensively researched transition metal oxides for decades. This is because the strong interplay between the lattice and the electronic (charge, spin, orbital) degrees of freedom leads to rich physics and novel electronic states \cite{Dagotto,Cheong,Tokura}. Owing to these competing degrees of freedom, small perturbations can significantly alter the physical properties of these oxides leading to emergent states of matter. A special class of transition metal oxides are the spinel vanadates with the AV$_2$O$_4$ structure. In this class of materials, the strongly correlated physics is accompanied by a delicate balance of geometrical frustration, orbital order, electron itinerancy, and competing magnetic interactions that lead to complex magnetic structures at low temperature \cite{Garlea,Kumar, Dalmau,Ma,Kismarahardja,Blanco,Kaur,SHLee,JHLee}.  They provide a unique playground for studying the strong interplay between spin frustration and the orbital degree of freedom.   

\noindent 

Spinel vanadates with a magnetic ion on the A-site (A = Fe, Mn) undergo multiple structural phase transitions, leading to low temperature non-collinear orbitally ordered ground states \cite{MacDougall,Kismarahardja2,KismarahardjaThesis,Kiswandhi,Nii,Suzuki,HDZhou,Adachi}. This complex behavior is due to competing exchange interactions ($J_{A-V}$ and $J_{V-V}$), spin-orbit coupling, and Jahn-Teller distortions on the V-site \cite{JHLee,Singha,motome}. In this class of materials, CoV$_2$O$_4$ is of interest due to its proximity to itinerancy from its short V-V distance; it is the closest known vanadate to an identified itinerant-localized crossover regime \cite{Kismarahardja,Kiswandhi}. Bulk CoV$_2$O$_4$ has been studied in powder and single crystal forms, showing orbital degeneracy lasting to very low temperatures. The material experiences a ferrimagnetic transition at T$_N$~=~150~K, with the \hkl<100> family of crystallographic directions being easy axes. Recently, a weak spin-canting and first order structural transition associated with an orbital glass transition has been identified at T~=~90~K \cite{Dalmau} (bulk samples). At this transition, the V-moments cant away from the Co-moment axis, at observed angles of about 5$^{\circ}$. Koborinai et al. \cite{Koborinai} showed a larger canting angle of 20\degree~in stoichiometric powder samples and 11\degree~for bulk single crystalline Co$_{1.3}$V$_{1.7}$O$_4$. Additionally, they reported temperature and field dependent magnetostriction effects, associating the canting with a lattice distortion in the material. Here we note, that based on the presented data in Koborinai et al. \cite{Koborinai}, the temperatures at which magnetic phase transitions occur as well as the severity of the magnetostriction, appear to be stoichiometry dependent. This shows that these spinel vanadates show extraordinary coupling between the spin, the lattice and the orbital degrees of freedom and that their magneto-elastic properties can be tuned via structural perturbations.  

The underlying mechanisms causing magnetostriction are the same as the mechanisms that cause anisotropy (e.g. crystal field, spin-orbit interaction, exchange interactions). In literature, mainly two mechanisms behind magnetostriction are discussed, i) tetragonal domain alignments and ii) spin-orbit coupling. Low-field magnetostriction ($B$ is smaller than the saturation field), is usually explained as alignment of tetragonal domains, which as the domain easy axis aligns with the magnetic field, leads to elongation of the crystal along the field and relative compression perpendicular to the field. This effect is reported in spinels such as MnV$_2$O$_4$\cite{Adachi,Suzuki,Suzuki_2009,Kismarahardja2,Liu2,Murikami} in which the onset of magnetic ordering is associated with a cubic to tetragonal distortion. However, in CoV$_2$O$_4$, magnetostriction is reported in the high temperature ferrimagnetic phase, which has a cubic structure, thus it is not clear that domain alignments are at play here. Koborinai et al. \cite{Koborinai} do not discuss the mechanism causing the observed magnetostriction. Even though Koborinai et al. \cite{Koborinai} claims a cubic structure (for 20$\%$ Co -rich single crystals) in the high temperature collinear ferrimagnetic state, their observed magnetostriction effects could be consistent with a domain picture at high temperature. It is clear that in bulk the magnetostrictive behavior changes at low temperatures when CoV$_2$O$_4$ undergoes a cubic to tetragonal phase transition. Here the system shows (short-range) orbital order and elongation perpendicular to the field with compression along the field direction. This low temperature behavior is consistent with the following explanation; as the magnetic field is turned on, the orbital occupation changes, resulting in distortions as the sample is elongated along the direction of the occupied orbital \cite{Kismarahardja2,Adachi,Suzuki_2009}.

Recently, we reported on the growth and characterization of CoV$_2$O$_4$ thin films grown on SrTiO$_3$ substrates, resulting in a stable orthorhombic structure even to film thicknesses of several hundred nanometers \cite{Christie}. We found that in this form, the material shows much stronger signs of spin canting and long-range orbital order than bulk, attributed to a symmetry lowering due to planar compressive strain, pushing the material further into the insulating state. From magnetization measurements and neutron diffraction in zero field, we determined the magnetic structure of the films. Below T$_N$~=~150~K, the material enters a ferrimagnetic state with an easy axis out of the plane, ($c$ = \hkl[001]), i.e., perpendicular magnetic anisotropy. This out-of-plane easy axis is a result of the stretched unit cell in that direction. 
At low temperature, we found a non-collinear regime, with the easy axis along the planar \hkl[110] direction and with a V-moment canting angle of at least $\sim$ 20$^{\circ}$. This major spin reorientation as a function of temperature is a sign that the anisotropy of the material changes drastically as the sample transitions into the spin-canting regime \cite{Christie}. This result has been reproduced recently by others \cite{Behera_2021}.

In this paper, we investigate the magnetic anisotropy of thin films of CoV$_2$O$_4$ grown on SrTiO$_3$ substrates through torque magnetometry measurements. From these measurements, we confirm the anisotropy change between the two magnetic regimes, and present a discussion of the strength of the different anisotropy states, including a very strong anisotropy in the low temperature state. In contrast to the first order structural phase transition seen in stoichiometric bulk powder samples at 90~K ($B$~=~0~T) \cite{Dalmau}, the observed magneto-crystalline anisotropy change occurs gradually over a rather large temperature range. This is reminiscent of the cubic-tetragonal second-order structural phase transition observed in SrTiO$_3$ \cite{Salman,Fanli}. Here we note, that Koborinai et al. \cite{Koborinai} report that a second order structural distortion is responsible for the observed temperature-dependent changes in their magnetostriction measurements. The nature of that distortion is not discussed and the investigated samples are Co$_{1.21}$V$_{1.79}$O$_4$ single crystals. We also find clear evidence of magnetostriction effects in the collinear ferrimagnetic state. Of further interest is the observation of planar \hkl<100> low temperature easy axes in torque rotations, different from the \hkl[110] easy axis seen in zero-field neutron results. We thus conclude that magnetostrictive behavior in the films leads to magnetic-field induced changes to the crystal structure and the measured anisotropy. 

\section{EXPERIMENTAL DETAILS}

High quality thin films of CoV$_2$O$_4$ have been grown onto (001) SrTiO$_3$ substrates via pulsed laser deposition using a home-made pressed pellet of CoV$_2$O$_6$. The films were grown in a background pressure of P = 1x10$^{-7}$ Torr, while the substrate was heated to a constant temperature ranging between 600 - 650 $^{\circ}C$. More details on the film growth and structural characterization have been previously published elsewhere \cite{Christie}. Atomic force microscopy (AFM) and x-ray diffraction (XRD) measurements were used to verify the thin film quality. All samples reported on in this manuscript have the orthorhombic structure with lattice parameters $a$~=~8.36(2)~\AA, $b$~=~8.24(5)~\AA, and $c$~=~8.457(3)~\AA 
 \cite{Christie}, rather than the bulk cubic structure with $a$=$b$=$c$=~8.407~\AA \cite{Dalmau}. While the films are epitaxial in the sense that the crystallographic directions are aligned with those of the substrate, the in plane lattice parameters are not lattice matched to the substrate lattice parameters (the mismatch is ~7$\%$). The in-plane lattice parameters are compressed by about 0.6$\%$ and 2$\%$ for the $a$ and $b$ directions, respectively.  In other words, by growing CoV$_2$O$_4$ on SrTiO$_3$, we stabilize a new orthorhombic phase, i.e., the films do not relax to the bulk cubic structure, even at large thicknesses of 400 nm, which we have already reported on before \cite{Christie}. 
 
 Information on the microstructure and the chemical composition are investigated using a probe-aberration-corrected cold field emission JEM-ARM200cF transmission electron microscope at 200 kV, which is equipped with Oxford energy dispersive spectroscopy (EDS) detector for composition analysis. The EDS results show that the Co:V stoichiometry is very close to the expected value of 1:2 (see supplemental materials for more details \cite{suppmat}). Furthermore, all measured samples show the same overall magnetization behavior compared to samples presented in the previously published neutron diffraction and magnetization measurements \cite{Christie}.  Magnetization measurements as a function of applied field taken at various temperatures
were performed in a Quantum Design magnetic properties measurement system (MPMS).

Torque magnetometry measurements were performed at the National High Magnetic Field laboratory (NHMFL) in fields up to 30~T. Measurements included both capacitive and piezoresistive torque magnetometry, performed in the NHMFL superconducting magnet SCM2 and 31~T resistive magnet Cell 7 (capacitive torque with samples mounted on CuBe cantilever, see Fig. \ref{RotationPlanes}a)), and a Quantum Design Physical Properties Measurement System (PPMS, piezoresistive torque with samples mounted on a Si substrate cantilever). Both mounts are able to rotate the sample along a plane. Interactions between the applied field and sample moment, and also between the easy axis and the sample moment, both generate a torque leading to deflection of the cantilever. This deflection is dependent on the magnetization of the sample and is measured as a function of applied field strength and direction (see Fig. \ref{RotationPlanes}a)). In capacitive torque magnetometry, the cantilever comprises the top plate of a parallel plate capacitor, and the deflection is measured as a change in capacitance, $\Delta C$, which is directly proportional to the torque $\tau$. An Andeen-Harling AH2700A Capacitance Bridge operating at frequencies between 1,000,294 Hz and 7,000 Hz was used to collect the capacitance data during each measurement. In piezoresistive torque magnetometry, the deflection is measured as a torsion response with calibrated piezoresistive elements, such that the torque value ($\tau$) is reported. The Tq-Mag option for the PPMS was used to measure the piezoresistive elements. In measurements shown in this paper, zero-field responses have been subtracted to remove background contributions such as gravity. The torque response for a SrTiO$_3$ substrate has been evaluated at 30~K at 9~T,  and it was found to be negligible compared to the response from the thin film (see the supplemental materials for more details \cite{suppmat}).

\section{RESULTS AND DISCUSSION}
\begin{figure}
    \centering
    \includegraphics[width = 3.41 in]{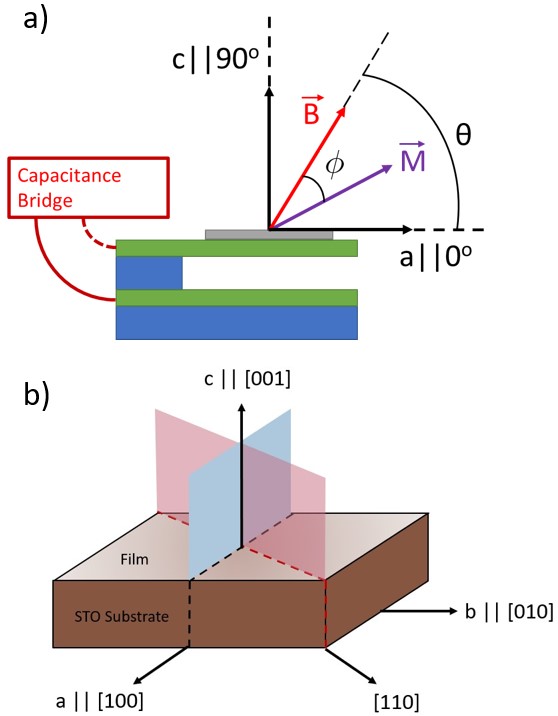}
\caption{a) Schematic diagram of the capacitive torque magnetometry set-up, which shows the sample placement on a flexible BeCu cantilever acting as the top plate of a parallel plate capacitor. The setup is rotated in magnetic field. b) Graphic showing the rotation planes. For this paper, the blue colored plane is referred to as the \hkl(010), the red as the \hkl(-110), and the light tan as the \hkl(001) rotation plane. 
    }
    \label{RotationPlanes}
\end{figure}

Torque measurements were performed at various temperatures, as a function of field direction, and as a function of the strength of the applied field. For the rotations, the change in capacitance $\Delta$C (or torque $\tau$) was measured as the sample was rotated in a constant applied field and at a constant temperature through three separate rotation planes colored in Fig. \ref{RotationPlanes}b). These planes are labeled by their surface normal: the \hkl(010) plane is colored blue, the \hkl(001) is colored light tan, and the \hkl(-110) plane is colored red. These measurements are aimed to explore the temperature and field evolution of the relative magnetic anisotropy of various crystallographic directions. Note, previous zero-field neutron measurements and magnetization measurements have shown that the zero-field low temperature easy axis is the \hkl[110] direction and the high temperature ferrimagnetic state easy axis is the \hkl[001] direction \cite{Christie}. In this paper we will show that the magneto-crystalline anisotropy changes dramatically as a function of temperature and applied magnetic field.

A discussion of some torque cases is useful to understand the CoV$_2$O$_4$ film measurements. In a torque magnetometry measurement, torque may be generated from two sources: a Zeeman energy term from the separation between the magnetization direction and the applied field, and an anisotropic energy term from the separation between the magnetization direction and the easy axis. If the magnetic moment is fixed along an easy axis due to the crystal field or intrinsic anisotropy, and not following the direction of the external B-field, the torque response will only reflect the Zeeman term $\tau~=~\vec{M}\times\vec{B}~=~|M||B|$sin$(\phi)$. Here the angle $\phi$ is the angle between the magnetization and the B-field. Zero torque occurs when the magnetic field is either parallel or anti-parallel to the magnetization.

In the other case, where the magnetic easy axes are weak and easily overcome by the magnetic field, the magnetization will remain saturated along the field direction through the entire rotation. Here the torque arises purely from the angular separation of the magnetization (i.e., the magnetic field) to the easy axes, or more precisely, the gradient of the magneto-crystalline anisotropy energy surface. Note, this energy surface is defined by the amount of energy it takes to rotate the magnetization of the sample along a rotational plane. Hence, the easy and hard axis of these rotational planes are defined where the magneto-crystalline anisotropy energy reaches its minima and maxima, as these axes require the lowest and highest energy to magnetically saturate, respectively. It follows that the torque is then given by the difference in anisotropic energy across an angular deviation, or its derivative ($\tau$ = $-\frac{dE}{d\theta}$). As a result, the torque reading for both easy and hard axes is zero.  The symmetry of the torque response for a single-crystalline material will reflect the symmetry of its unit cell. For a cubic unit cell, a torque rotation over two easy axes separated by 90$\degree$ results in a $\sin(4\theta)$ response assuming the system remains magnetically saturated with the field. In the case of a tetragonal distortion where one of the easy axes is lost with the breaking of cubic symmetry, torque will show an uniaxial response of $\sin(2\theta)$.  \cite{Cullity,Handley}

\begin{figure}
    \centering
    \includegraphics[width = 3.41 in]{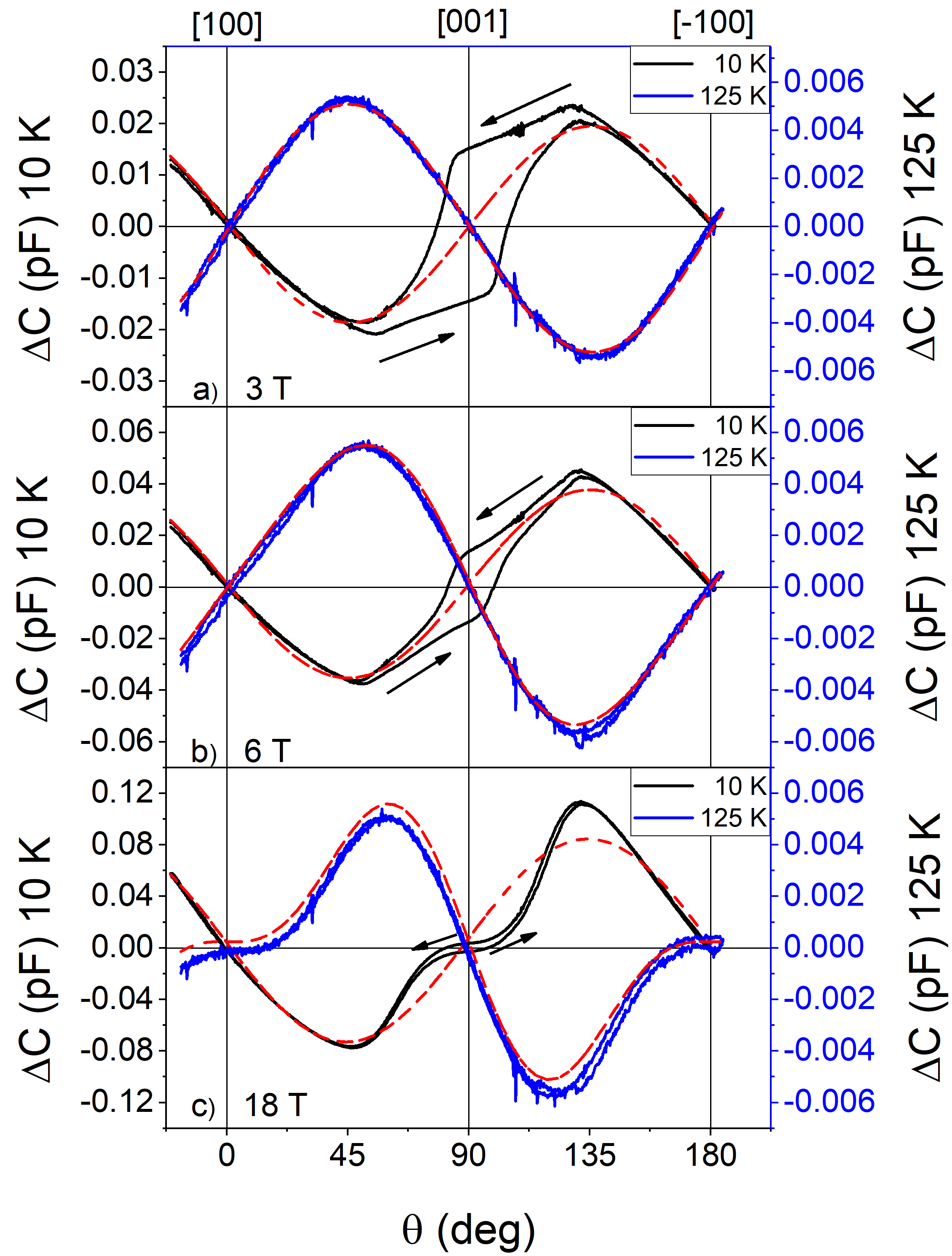}
    \caption{\hkl(010) rotation curves for a 260~nm thick film in B-fields of a) 3~T, b) 6~T, and c) 18~T. Rotations were performed at 10~K (black solid curves, left axis) and 125~K (blue solid curves, right axis). Red dashed lines (10~K curves) are shown for reference and represent perfect uniaxial responses, while the red dashed lines (125~K curves) are fitted lines for a combined uniaxial and biaxial model (see text). Vertical black lines are added to indicate crystal axis directions. 
    }
    \label{OOPRotationsFit}
\end{figure}

Fig. \ref{OOPRotationsFit} shows representative \hkl(010) torque rotation curves taken on a 260~nm thin film sample, plotted as the change in capacitance $\Delta C$ as a function of the angle $\theta$, which is defined as the angle between the applied field $\vec{B}$ and the \hkl[100] direction (see Fig. \ref{RotationPlanes}b)). Rotations taken with $B$= 3, 6, and 18~T are shown for two temperatures, at T = 125~K (ferrimagnetic state, blue curves) and at $T$ = 10~K (non-collinear state, black curves). 
In the high temperature ferrimagnetic regime, the torque shows an almost perfect uniaxial $\sin(2\theta)$ behavior at 3~T. A notable feature of this set of rotations is the approximately constant amplitude of the capacitance change for rotations with $B \geq$ 3~T, indicating that the magnetization is saturated along the field direction through the entire rotation plane at all fields shown in Fig. \ref{OOPRotationsFit}. 
Thus, in the high temperature collinear ferrimagnetic state, the torque is determined only by the magneto-crystalline anisotropy that is dictated by the shape of the unit-cell. For a cubic unit cell the response would be biaxial, however in our thin films the unit cell is elongated along the out-of-plane direction of the film (compared to the in-plane lattice parameters). Thus, in the high temperature collinear state the torque confirms that the elongated $c$-axis is the easy axis (a negative slope around the zero crossing associated with the \hkl[001]) at 125~K \cite{Cullity}. When inspecting the variation from low to high fields, we see the torque maximum and minimum shift closer to the out-of-plane direction causing the slope to increase with the field applied near the easy axis, and flatten out near the hard axis. In other words, while the overall torque amplitude remains the same, there is a growing $\sin(4\theta)$ contribution to the response as the field increases. 

\begin{figure}
    \centering
    \includegraphics[width = 3.41 in]{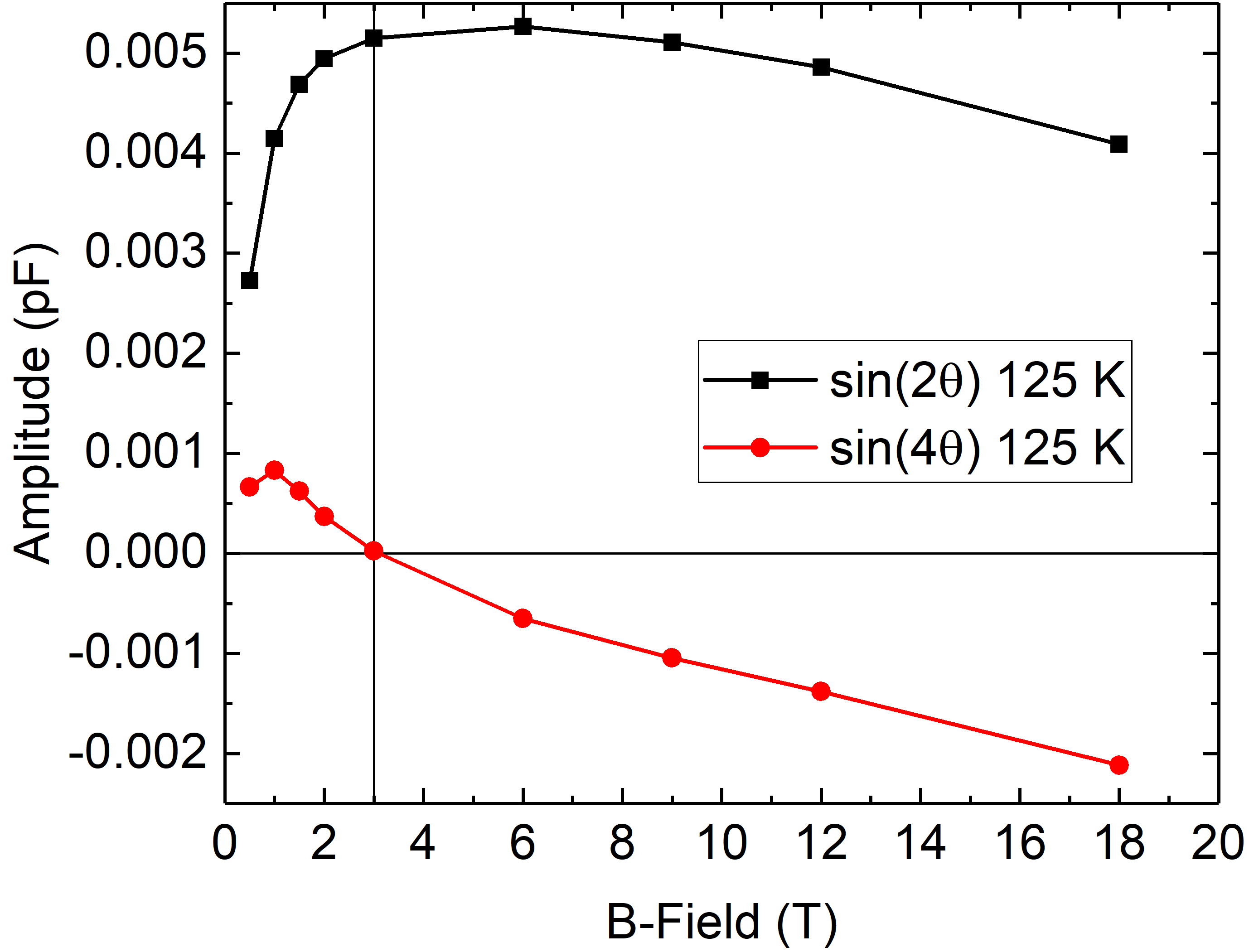}
    \caption{Plot of sinusoidal amplitudes of fitted curves for a 260~nm thick sample at 125~K for \hkl(010) rotations. The torque curves are fitted to a A$_1$sin(2$\theta$) + A$_2$sin(4$\theta$) equation. 
    }
    \label{AnisoConstvsH}
\end{figure}

The 125~K field dependent behavior shown in Fig. \ref{OOPRotationsFit} is summarised in Fig \ref{AnisoConstvsH}, where the torque curves are fitted to A$_1$sin(2$\theta$) + A$_2$sin(4$\theta$) and the amplitudes of the sinusoidal fits are shown as a function of the applied magnetic field. At 3~T, the anisotropy shows a near perfect uniaxial fit, as noted before. Above 3~T, the negative biaxial component of sin(4$\theta$) replaces the uniaxial sin(2$\theta$) component such that the difference in amplitude is constant. The constant difference is representative of the magnetic saturation, so the sum of absolute values of the uniaxial and biaxial component is field independent. But below 3~T, there is a sharp decrease in sin(2$\theta$) and the smaller, positive increase in sin(4$\theta$). So, the overall amplitude of the torque curves decreases with field, an indication that below 3~T the torque curve is unsaturated. The low-field torque responses and their corresponding fits are provided in the Supplemental Materials \cite{suppmat}. 
The introduction of a biaxial term to describe unsaturated torque curves of an uniaxial system is in line with several works published by others \cite{Ono1,Handley,Margulies,Ono2}. 
These observations all support the following picture: once the system reaches saturation ($B\geq$3~T), the symmetry of the torque response starts to gradually change from sin(2$\theta$) to sin(4$\theta$). Hence, the anisotropic energy surface near the in-plane hard axis is changing, effectively softening this axis and trending towards a biaxial response, i.e., magnetostriction leads to a distortion of the unit cell, making it more cubic. Extrapolating the behavior seen in Fig. \ref{AnisoConstvsH}, we find that the structure should become fully cubic (i.e., a pure biaxial response) at around $B$=59.35 $\pm$ 6.21~T.

For off-stoichiometric bulk single crystals \cite{Koborinai}, low field magnetostriction in the collinear ferrimagnetic state reportedly shows that the unit cell distorts away from cubic symmetry. In thin films, we find that high field magnetostriction makes the magnetic anisotropy, ergo the unit-cell structure, more cubic. It is possible that similar to one report on MnV$_2$O$_4$ \cite{Liu2}, the film breaks up into multiple domains at high field, which would make the sample appear more isotropic compared to a single tetragonal-like domain scenario at lower fields. Another possible scenario is that field induced changes to the orbital occupation lead to the observed changes in anisotropy.

In the low temperature non-collinear regime the \hkl(010) rotation plane shows a more complicated response than for the ferrimagnetic regime. One thing is very clear, the torque response has flipped sign in amplitude, meaning the uniaxial easy axis has changed from the \hkl[001] to the in-plane \hkl[100] direction. Based on our previous work \cite{Christie}, we found no evidence of a temperature-induced change in lattice parameters as the sample was cooled through the 90~K transition. As discussed below, we associated this anisotropy change with changes in oxygen positions around the V-site, i.e., at low temperature, the local single ion anisotropy dominates compared to the shape of the unit-cell. In the Supplemental Materials \cite{suppmat} we provide the torque response for the \hkl(-110) plane, which shows that relative to the out-of-plane direction, the \hkl[110] direction is also easy. 
We will discuss the relative anisotropy of the in-plane directions in more detail below.  An asymmetry of the torque amplitude is observed in higher field: this is likely due to nonlinear behavior of the cantilever flexing up or down as these larger torques are applied to the cantilever. While the uniaxial trend with a planar easy axis is clear, hysteresis around the \hkl[001] axis indicates a more complex balance of the Zeeman and anisotropic torque terms around the hard direction. This hysteresis occurs as the field is not large enough to saturate the magnetization along that hard axis \cite{Lisfi_2014,Lisfi_CFO}, i.e., the applied field is smaller than the anisotropy field. The magnetization does not align with the field when the field points along the \hkl[001] direction at low temperature. As the magnetization lags behind the applied field while rotating through this hard axis, a Zeeman energy term results in non-zero torque along the hard axis and hysteresis depending on the direction of the sample rotation. 
We visualize this hysteresis by plotting the difference in torque between the up and the down sweep, which is plotted in Fig. \ref{TorquevsH} a). Utilizing the in-plane saturation magnetization (see Supplemental Materials \cite{suppmat} for more details on MPMS measurements), we can estimate the maximum angle $\phi$ between $\vec{M}$ and $\vec{B}$ when the field is parallel to the hard axis. The Zeeman angle is largest when $\vec{B}||$\hkl[001] and decreases with increasing applied field (see Fig. \ref{TorquevsH}b)). We warn that these Zeeman angles should be considered estimates, and only give a general idea for angular separation between $\vec{M}$ and $\vec{B}$. The exponential decrease of the Zeeman angle with increasing field indicates that the anisotropy field significantly exceeds 10~T in the low temperature noncollinear phase. The hysteresis slowly closes as the applied field reaches 18~T.

The lack of saturation in the non-collinear phase was further examined through torque field sweeps ($\Delta C$ vs. $B$), where the field was swept between $\pm 18$ T. The field sweep of the non-collinear phase along the hard \hkl[001] axis results in a butterfly loop (Fig. \ref{TorquevsH}c)). Note that if magnetization is saturated in the field direction, then a torque response at a hard axis, halfway between two easy axes, should be zero; a saturated field sweep at such an axis should result in zero response as a function of the applied field. Instead, the hysteretic butterfly pattern indicates the magnetization deviates from the field direction generating a nonzero torque even at high fields. Arrows in the figure indicate the direction of the field sweep when sweeping between -18~T to +18~T. At around $\pm$ 13~T, the hysteresis closes, but the torque remains nonzero.
Upon reaching this coercive field, the torque response flips signs when ramping back to zero. This could be due to a small misalignment of the field with respect to the \hkl[001] direction, which would be consistent with the small residual torque above the coercive field.  
 
It is also possible that the high fields alter the magneto-crystalline energy surface and that a magnetostrictive distortion of the lattice contributes to the torque signal, which could be at least in part responsible for the lattice favoring the opposite easy axis on the down sweep. The presence of magnetostriction is evident in the high temperature state as described above, furthermore, a biaxial term also appears to develop at low temperature as a function of increasing field, similar to what is observed in the collinear ferrimagnetic regime. As before, the torque curve flattens near the hard axis with increasing applied fields (see Fig. \ref{OOPRotationsFit}c)). The hysteresis makes it hard to fit the torque curves in the non-collinear phase, however, the curves at high field can be modeled using an uniaxial and a biaxial anisotropy term (see the Supplemental Materials \cite{suppmat} for more details). The high field model curves show that the biaxial contribution appears to develop similarly to what we plotted for the high temperature state in Fig. \ref{AnisoConstvsH}, but with both anisotropy constants showing negative values. This is consistent with the change in unaxial easy axis direction of 90\degree, which affects the sign of the uniaxial term but not of the biaxial term. While the presence of hysteresis and the development of a biaxial term in applied field is very reproducible from sample to sample, how quickly the magnetostriction develops with applied field strength does appear to be sample dependent (see Supplemental Materials \cite{suppmat} for more details). This sample to sample variation may be due to subtle microstructural differences between the samples. %

\begin{figure}
    \centering
    \includegraphics[width = 3.41 in]{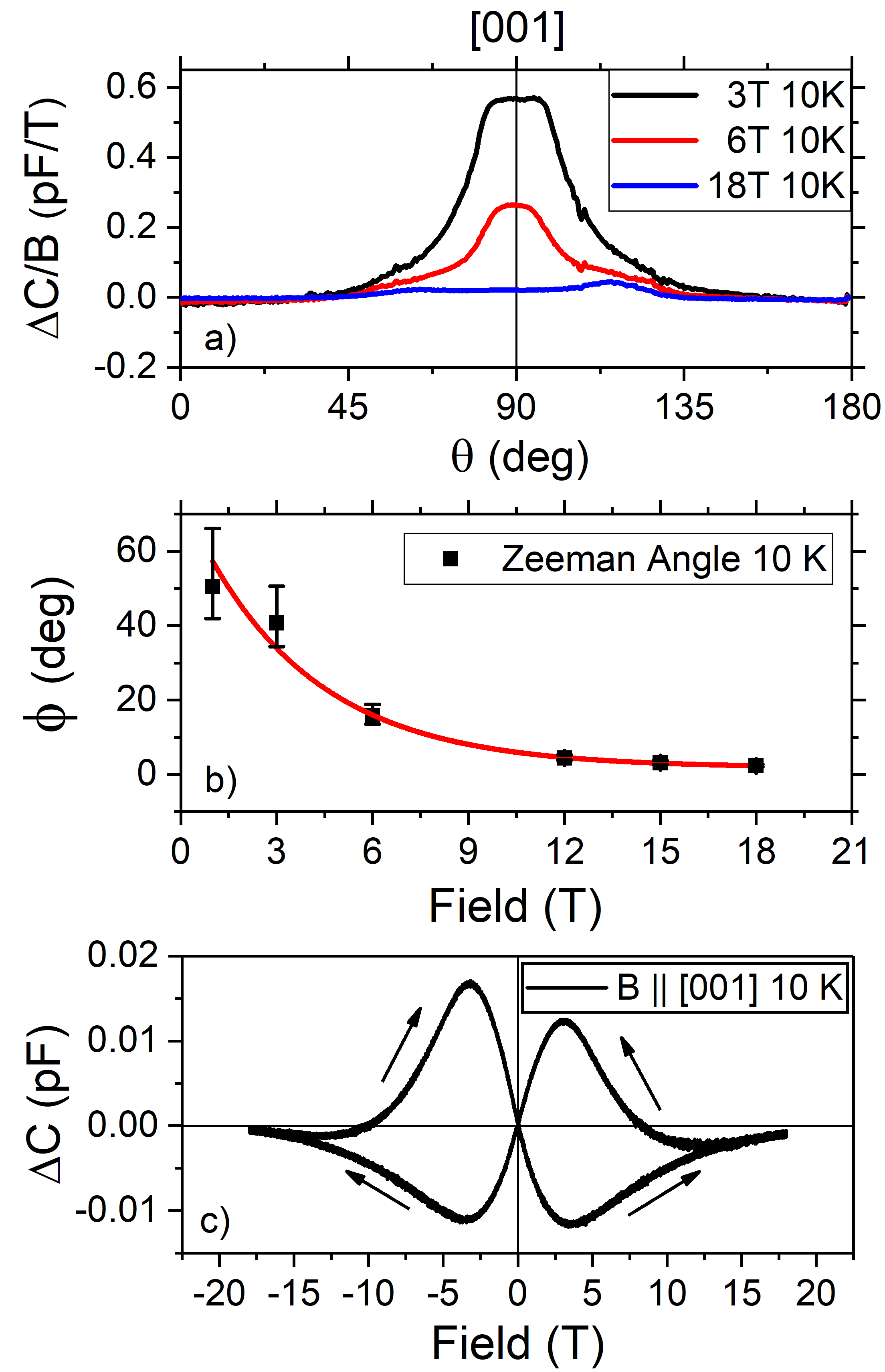}
    \caption{a) Field-scaled hysteresis for rotations shown in Fig. \ref{OOPRotationsFit} for applied fields of 3~T, 6~T, and 18~T plotted as a function of $\theta$. b) Calculated Zeeman angle amplitude as a function of applied field ($\vec{B} ||$ \hkl[001]) with magnetic moment from MPMS measurements (see Supplemental Materials \cite{suppmat}) taken as 9.19 $\mu_B$ per unit cell. Error bars are from error propagation of the uncertainty in the sample volume (2.81 $\pm$ 0.43) $\times$ $10^{-13}$ m$^{3}$ used to determine the magnetization and the Zeeman Angle. The red line is an exponential fit. c) Field sweeps of sample torque response $\Delta$C vs. $B$, with field applied along the sample's \hkl[001] axis, in the low temperature non-collinear phase. 
    }
    \label{TorquevsH}
\end{figure}

\begin{figure}
    \centering
    \includegraphics[width=3.41in]{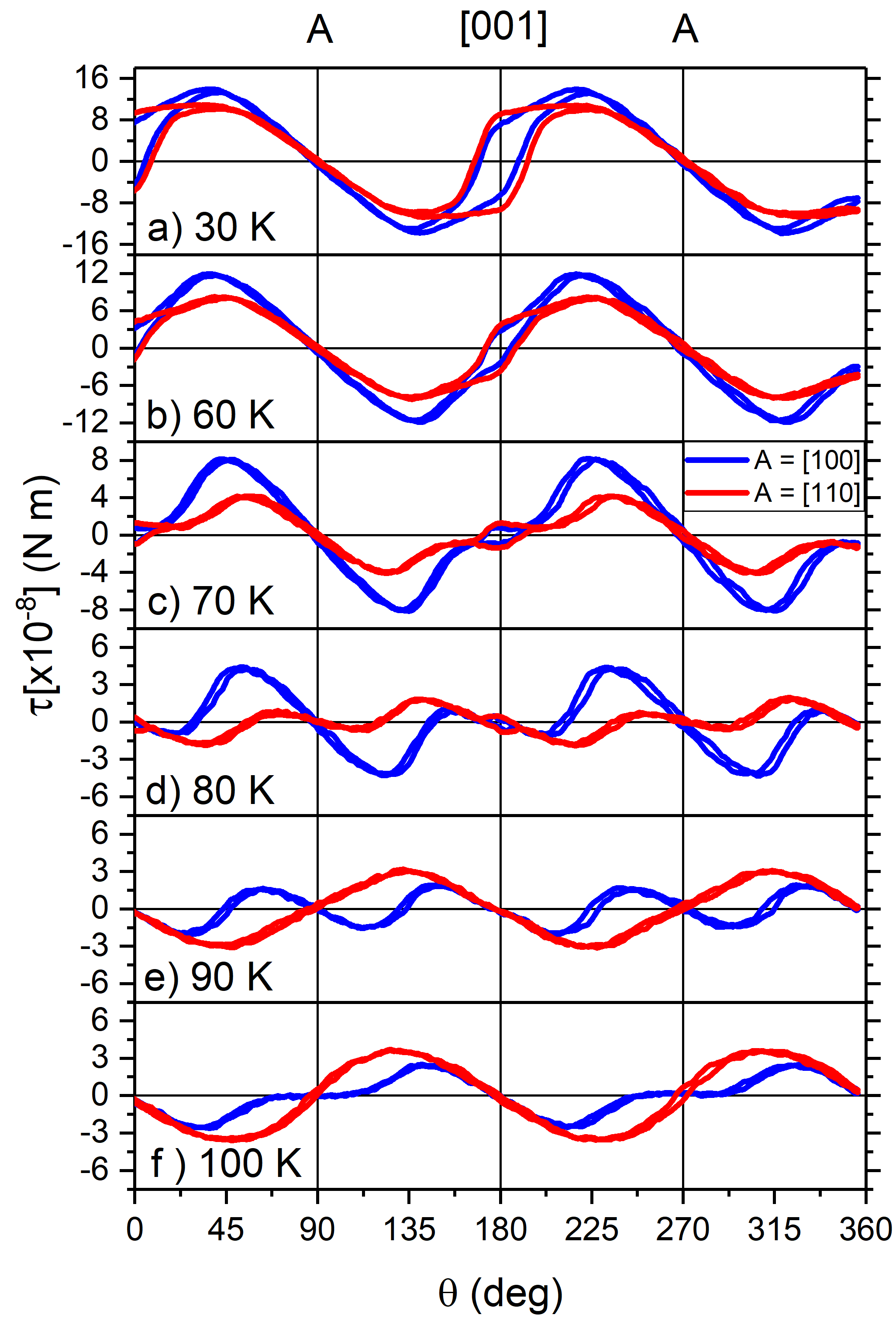}
    \caption{Torque rotations (260~nm thin film) for two rotation planes (shown in red: \hkl(-110) and shown in blue: \hkl(010)) measured at 6~T at temperatures of 30~K, 60~K, 70~K, 80~K, 90~K, and 100~K. Vertical gray line labeled \hkl[001] represents when the magnetic field points out-of-plane. The letter ``A" represents when the magnetic field points along the planar \hkl<110> and \hkl<100> directions for the red and blue curves, respectively. 
    }
    \label{OOPMidT}
\end{figure}

To further explore the relative anisotropy of the \hkl[100], \hkl[110], and \hkl[001] directions, additional torque measurements were done from 30~K to 100~K at 10~K increments for two rotational planes: \hkl(010) and \hkl(-110). Full rotations in a magnetic field of 6~T are shown in Figure \ref{OOPMidT}.
In this graph we compare the temperature evolution of the torque response to highlight the changes in relative anisotropy between the \hkl[001] and the planar \hkl[100] and \hkl[110] directions. The in-plane directions for each rotation is marked with the letter ``A" (at 90$\degree$ and 270$\degree$, respectively ). Starting from 30~K, the \hkl(010) (blue curve) rotation displays a larger amplitude than the \hkl(-110) (red curve) rotation, suggesting that at this temperature at $B$~=~6~T, the magnetization would favor alignment with the \hkl[100] planar directions rather than the \hkl[110] planar directions. Furthermore, a clear hysteresis loop forms near 0\degree  and 180\degree, characteristic of the \hkl[001] hard axis. Due to this hysteresis at the edge of our rotation, the curves at 0$\degree$ and 360$\degree$ do not overlap. This is because we reverse rotation before completing the hysteresis loop, i.e., the sample is in a domain state when we reverse direction; at 0$\degree$ (360$\degree$) the sample resides on the upper (lower) branch of the hysteresis loop. With increasing temperature, the torque amplitude difference between the red and the blue curves increases. At 70~K, the \hkl(010) (blue curve) rotation displays an amplitude roughly twice that of the \hkl(-110) (red curve) rotation. For both planes, the response flattens out around the \hkl[001] direction eventually leading to the slope around the \hkl[001] flipping signs as the sample is heated to 80~K, while maintaining a small amount of hysteresis. Important to note is that at 80~K the response in the \hkl(-110) plane shows a mostly $\sin(4\theta)$ symmetry indicating the formation of a domain state with equal volume fractions of material for which the \hkl[110] and \hkl[001] directions are easy, respectively. In comparison, the \hkl(010) rotation at this temperature still maintains most of its original $\sin(2\theta)$ pattern. At a higher temperature of 90~K, the \hkl(010) plane shows a mostly $\sin(4\theta)$ response, indicating the equivalence of the cubic directions in the crystal. At this higher temperature the planar \hkl<110> directions are already hard axes compared to the out-of-plane direction, signaling the onset of the uniaxial anisotropy $\sin(2\theta)$ of the collinear ferrimagnetic state. Finally, at 100~K, the \hkl(010) plane also shows the onset of uniaxial response, but with a clear $\sin(4\theta)$ contribution remaining. This gradual change in magneto-crystalline anisotropy is consistent with a second-order phase transition, which develops in our torque measurement as a continuous evolution of a domain state over a broad range of temperatures between 30~K to 100~K. This is in stark contrast with observations of a first-order structural transition seen in stoichiometric bulk CoV$_2$O$_4$ at 90~K (with $B$~=~0~T) \cite{Dalmau}. As noted before, Koborinai et al. \cite{Koborinai} report that a second order structural distortion is responsible for the observed temperature-dependent changes in their magnetostriction measurements. Our second-order transition is reminiscent of the 105~K transition in SrTiO$_3$, in which adjacent TiO$_6$ octahedra rotate, lowering the crystal symmetry from cubic to tetragonal. We draw the analogy with observations in SrTiO$_3$ \cite{Salman, Fanli} because in CoV$_2$O$_4$ thin films we also observe that the symmetry lowered phase nucleates over a rather large temperature range, likely in a spatially inhomogeneous way. 
Indeed, a recently published theoretical work by Sharma et al. \cite{Sharma-DFT}, indicates that distortions of the VO$_6$ octahedra drive the spin reorientation in this spinel vanadate. 

These torque measurements show that the planar \hkl<100> directions are the easy axes at all temperatures in applied fields of 6~T. This appears to be in contrast with our previously published zero-field easy axis of \hkl[110] based on \emph{zero-field} neutron experiments \cite{Christie}. To address this discrepancy, we show the temperature evolution of the same rotation curves as in Fig. \ref{OOPMidT} but for fields of 1.5 and 3~T in the supplemental materials \cite{suppmat}. These additional measurements show that while the overall temperature dependence of these curves is independent of applied field strength, as the field is reduced, the planar \hkl<100> and \hkl<110> directions become more equivalent, i.e., the \hkl(010) and \hkl(-110) torque rotations have the same low temperature amplitudes. This indicates that as the field is lowered the low temperature difference in relative anisotropy between the \hkl[100] and \hkl[110] directions diminishes. Extrapolating, this shows that the torque measurements are not in disagreement with the zero-field neutron measurements, rather the magneto-crystalline anisotropy is very sensitive to applied field due to substantial magnetostriction at low field, which is in line with reports by others \cite{Koborinai} on bulk samples. At low temperature, it is expected that the field induces a different distortion as the anisotropy is now more closely tied to the local structure around the individual Co and V sites (i.e., orbital occupation), rather than the shape of the unit cell. A change in orbital occupation in applied field would explain why our easy axis is along the planar \hkl<100> directions in applied field opposed to the \hkl[110] easy axis, which we measured in zero-field neutron measurements before.  

\begin{figure}
    \centering
    \includegraphics[width = 3.42 in]{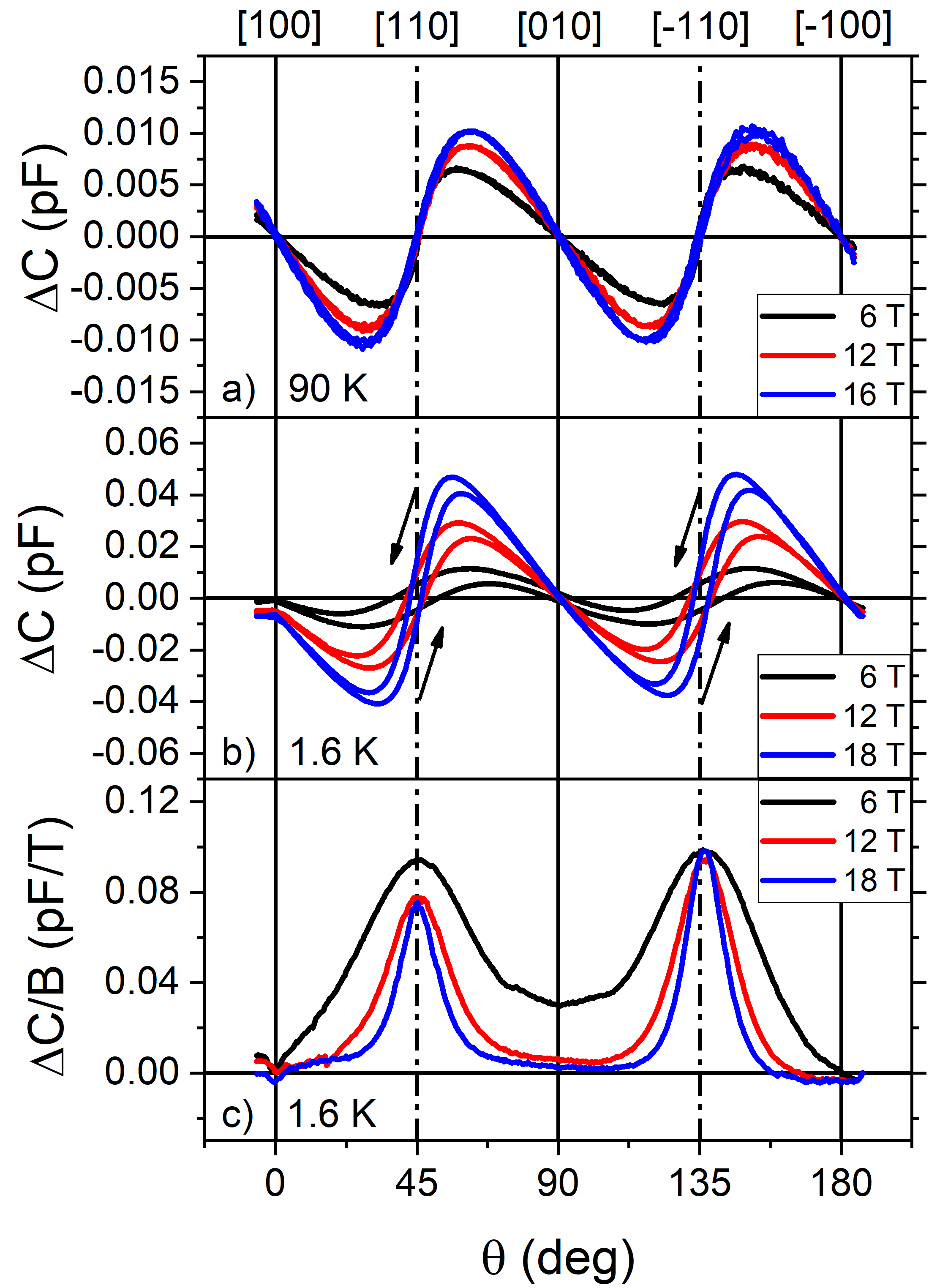}
    \caption{Torque rotations (140 nm thin film) for the \hkl(001) plane. a) Rotations performed at 90~K in B = 6, 12, 16~T and b) at 1.6~K in B = 6, 12, 18~T. Arrows indicate the direction of the rotation for hysteretic portions of the curves. c) Field-scaled hysteresis extracted from the rotations presented in panel b. The solid and dashed black vertical lines are associated with the cubic and \hkl<110> planar crystallographic axes, respectively.  
    }
    \label{IPRotations}
\end{figure}

Next, we directly compare the relative anisotropy of the planar directions as a function of temperature and applied field. Representative torque rotations for the \hkl(001) rotation plane are shown in  Fig. \ref{IPRotations}, taken in fields ranging from 3~T to 18~T at 1.6~K and 90~K. 
An overall $\sin(4\theta)$ response is visible in both the ferrimagnetic and the non-collinear temperature regime, indicating a biaxial anisotropy. 

In the ferrimagnetic regime measured at 90~K, the overall easy axis in the \hkl[001] direction is not included in this rotation plane. However, we can measure the relative anisotropy of the planar \hkl<100> and \hkl<110> directions. Among the directions that lie on the \hkl(001) plane, the easiest axes are along the \hkl<100> directions. While the response shows a clear biaxial shape, it is distorted away from a perfect $\sin(4\theta)$, with a much steeper torque change around the \hkl<110> directions. Note, at 125~K the torque response is purely sinusoidal as expected for a purely biaxial system (see Supplemental Materials \cite{suppmat} for more details). Additionally, the torque amplitude increases with increasing applied fields, this shows that with more field we are gradually increasing the moment that is pulled away from the \hkl[001] easy axis into the film plane. All of this shows that the \hkl<110> planar axes are hard axes compared to the \hkl<100> planar axes in substantial applied fields, which is consistent with the rotations presented in Fig. \ref{OOPMidT}. 

In the non-collinear low temperature state, the \hkl(001) rotations still show an overall biaxial $\sin(4\theta)$ response, indicating that this state also has planar \hkl<100> easy axes. Additionally, the curves show hysteresis over the entire rotation range at lower fields, so the magnetization is always lagging behind the field direction. The hysteresis is most prevalent, and remains up to the highest measurement field, around the planar \hkl<110> directions. At low fields, the torque begins to lose its sinusoidal shape, this is likely due to the fact that the torque signal becomes too small to measure on top of the hysteresis. Other noteworthy features are the dramatic change in amplitude with increasing field and the distortion of the curve from a lower field sinusoidal response to a sawtooth shape at higher fields. While this behavior may look somewhat similar to what is seen in the ferrimagnetic regime, it is in fact dramatically different. It is important to remind the reader that at low temperature the easy axis lies in the plane of the film. While at low field the moment is easily saturated along the \hkl[110] direction as we reported before, the field dependent torque rotations show that the magnetic moment is not saturating along any planar direction, as evidenced by the field-dependent change in the torque amplitude. While at low fields there is no clearly definable easy axis, an increasing torque amplitude and the closing of the hysteresis around the \hkl<100> planar directions in high fields, points to the \hkl<100> planar directions as the easy axes in fields of 3~T or more. The continued hysteresis around the \hkl<110> planar directions and the increased sawtooth shape show that the planar anisotropy in the sample is rather large. This is emphasized by plotting the hysteresis for the low temperature rotations as a function of applied field strength (see Fig. \ref{IPRotations}c)). The width of the hysteresis peak decreases with field, such that the full width half maximum of the hysteresis exponentially decays with a power coefficient of -0.14 $\pm$ 0.01 T$^{-1}$ (see Supplemental Materials). Note, that magnetization measurements show fully saturated moments along the \hkl[100] and \hkl[110] directions at fields of the order of 2~T. However, in rotation experiments the moment will not saturate along the planar \hkl<110> direction until $\sim$ 30~T (extrapolated from the exponential decay).  This points to dramatic magneto-elastic effects dominating our rotation experiment, i.e., the magneto-crystalline anisotropy changes in even moderate applied fields. This is not surprising as substantial magneto-elastic effects have been reported before in bulk samples in fields of only $\sim$0.2~T \cite{Koborinai}. Here we note, that the temperature and field evolution of the hysteresis seen in the planar experiments (Fig. \ref{IPRotations}) varies somewhat from sample to sample. In the Supplemental Materials \cite{suppmat} we show a comparison between \hkl(001) rotations of the 140 and 260~nm thick samples, these results show that the behavior of the thinner sample (at 1.6~K) is comparable to the behavior in the thicker sample but taken at 30~K. This indicates that the anisotropy is less strong in the thinner sample and that relative to the other samples in this study, this sample has a stronger magnetostrictive effect as a function of applied field.



\section{CONCLUSIONS}
We have characterized the magneto-crystalline anisotropy of CoV$_2$O$_4$ thin films deposited on \hkl[001] oriented SrTiO$_3$ as a function of temperature and applied magnetic field. We confirm the previously demonstrated anisotropy change from a perpendicular to a planar easy axis as the sample is cooled. Interestingly, we find that the anisotropy change occurs in a second-order structural phase transition that spans tens of Kelvins. In accordance with ref. \cite{Sharma-DFT}, we speculate that this transition involves the distortion of the oxygen cages around the transition metal sites. This is in clear contrast with the first-order structural phase transition reported in bulk \cite{Dalmau}, but it is consistent with the second-order effects reported by Koborinai et al. \cite{Koborinai}. Furthermore, the low temperature state has a rather large magneto-crystalline anisotropy, i.e., fields of the order of 18~T are needed to fully align the magnetic moment with the out-of-plane hard axis and fields in excess of 30~T appear to be needed to align the magnetization with the in-plane \hkl<110> hard directions.  

Here we note that previous zero-field neutron measurements showed the easy axis to be along the \hkl[110] axis at low temperature, which is in line with magnetization measurements (see Supplemental Materials \cite{suppmat}). However, the torque rotations clearly indicate the planar \hkl<100> directions to be the easy axes at all applied fields. To address this discrepancy, we note that at all temperatures, applied fields distort the uniaxial response with an increasing biaxial term. This indicates that magnetostrictive effects are substantial in these thin films even at moderate applied fields, thus the applied field significantly alters the magneto-crystalline anisotropy. In line with reports by others, we attribute the field-induced change in the low temperature planar easy axis to lattice distortions associated with changed orbital occupancies \cite{Adachi,Suzuki_2009}. 

The exact mechanism of these large anisotropy changes is not yet understood, and requires additional investigation into the microscopic mechanisms behind the observed phase transitions, i.e., the study of orbital occupation and site-specific oxygen distortions become important topics of study. As noted in the supplemental section, although sample to sample variations are minute, the severity of the anisotropy and the magnetostrictive effects seem to vary between samples. We speculate that the magnetostriction effect depends on the stoichiometry of the samples, allowing for some samples to have higher magnetostrictive responses than others. Additional investigation will be required regarding the exact stoichiometry and composition of these samples, but their general magnetic characteristics remain the same. 


C.B. acknowledges support from the National Research Foundation, under grant NSF DMR-1847887. We acknowledge the user-support from David E. Graf with the torque magnetometry measurements. This work was performed at the National High Magnetic Field Laboratory, which is supported by National Science Foundation Cooperative Agreement No. DMR-1157490, No. DMR-1644779, and the State of Florida.

\bibliography{references.bib}

\end{document}


\title{Supplemental Materials: Large Magnetic Anisotropy and Magnetostriction in Thin Films of CoV$_2$O$_4$    }

  \author{Sangsoo Kim}\altaffiliation{These authors contributed equally to this work}
 \affiliation{National High Magnetic Field Laboratory, Florida State University, Tallahassee, FL 32310, USA}
  \affiliation{Department of Physics, Florida State University, Tallahassee, FL 32306, USA}
  \author{Christie Thompson}\altaffiliation{These authors contributed equally to this work}
  \affiliation{National High Magnetic Field Laboratory, Florida State University, Tallahassee, FL 32310, USA}
 \affiliation{Materials Science and Engineering Program, Florida State University, Tallahassee, FL 32310, USA}
 \author{Yan Xin}
  \affiliation{National High Magnetic Field Laboratory, Florida State University, Tallahassee, FL 32310, USA}
 \author{Christianne Beekman}\email{Corresponding author: beekman@magnet.fsu.edu}
 \affiliation{National High Magnetic Field Laboratory, Florida State University, Tallahassee, FL 32310, USA}
  \affiliation{Department of Physics, Florida State University, Tallahassee, FL 32306, USA}
\date{\today}

\maketitle
\renewcommand{\thefigure}{S\arabic{figure}}
Fig. \ref{TEM}(a) and (b) show representative high-angle annular dark field scanning transmission electron microscope (HAADF-STEM) images of a $\sim$175 nm CoV$_2$O$_4$ film grown on a \hkl[001] SrTiO$_4$ substrate. The film shows a sharp film/substrate interface. The diffraction pattern shown in Fig. \ref{TEM}(b), confirms that the lattice directions in the film are aligned with those of the substrate. The film is a single crystal with the structure of CoV$_2$O$_4$. Energy dispersive X-ray spectroscopy (EDS) from the film confirms that the film has a Co:V ratio of 1:2 with a slight Co deficiency (Fig. \ref{TEM}(d)).
The V atomic columns in the HAADF-STEM image show brighter intensity than the Co atomic columns (Fig. \ref{TEM} (c)), which is counter-intuitive, considering the atomic number $Z$ is bigger for Co ($Z$ = 27) than for V ($Z$ = 23). However, in this structure, along \hkl[100], the Co atomic columns contain only Co, while in the V columns, there are two additional oxygen atoms in between the V atoms.  The atomic column intensity in HAADF-STEM image is proportional to the atomic number $Z^2$, and the number of atoms in that column. The higher average atomic number per unit length, the higher the column intensity is. In this case, the Co column has an average $Z$ number of 3.21~\AA~ and V/O column has a 4.63~\AA. So the V/O columns have higher intensity, which is not surprising. This is also confirmed by image simulation (Fig. \ref{TEM}(c) inset). The simulation has shown the V/O columns are brighter than Co columns at a wide range of sample thickness.  These data confirm that the film is CoV$_2$O$_4$. 

\begin{figure}
    \centering
    \includegraphics[width = 5 in]{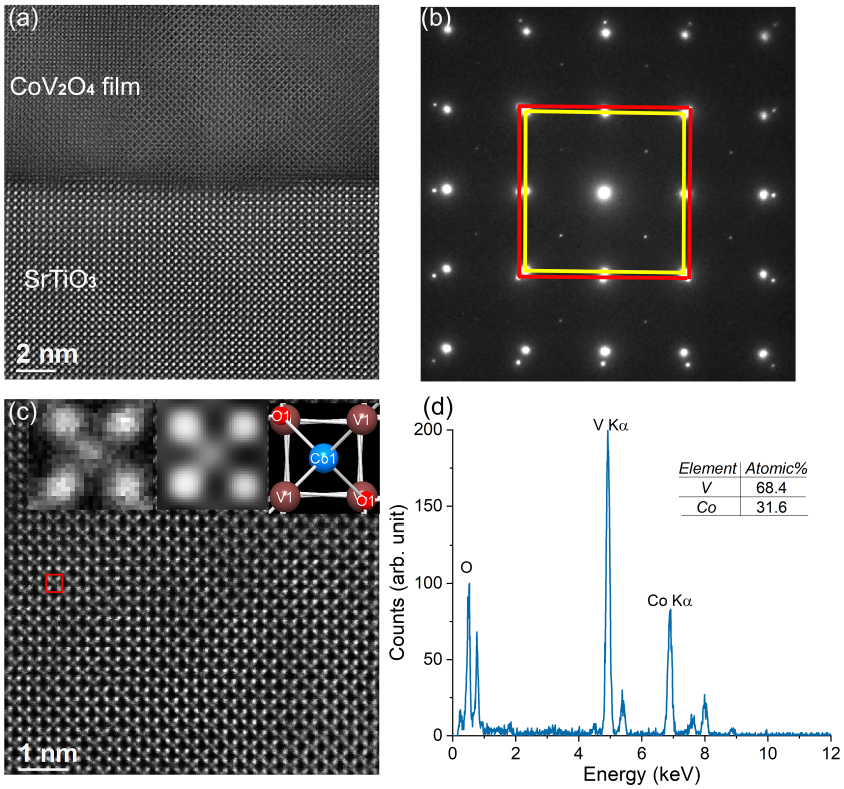}
    \caption{(a) High resolution HAADF-STEM image of the film-substrate interface viewed along \hkl[100].  (b) \hkl[100] Diffraction pattern including film and substrate.  The diffraction spots from the substrate are indicated by the red square and the ones from the film by the yellow square. (c) \hkl[100] atomic resolution HAADF-STEM image of the CoV$_2$O$_4$ film. Insets: blow-up of the red square from the experimental image (left); simulation at a thickness of 56 nm (middle); \hkl[100] projected view of Co and V atom relative positions in CoV$_2$O$_4$ structure. (d) EDS spectrum from the film.} 
    \label{TEM}
\end{figure}

\begin{figure}
    \centering
    \includegraphics[width = 7 in]{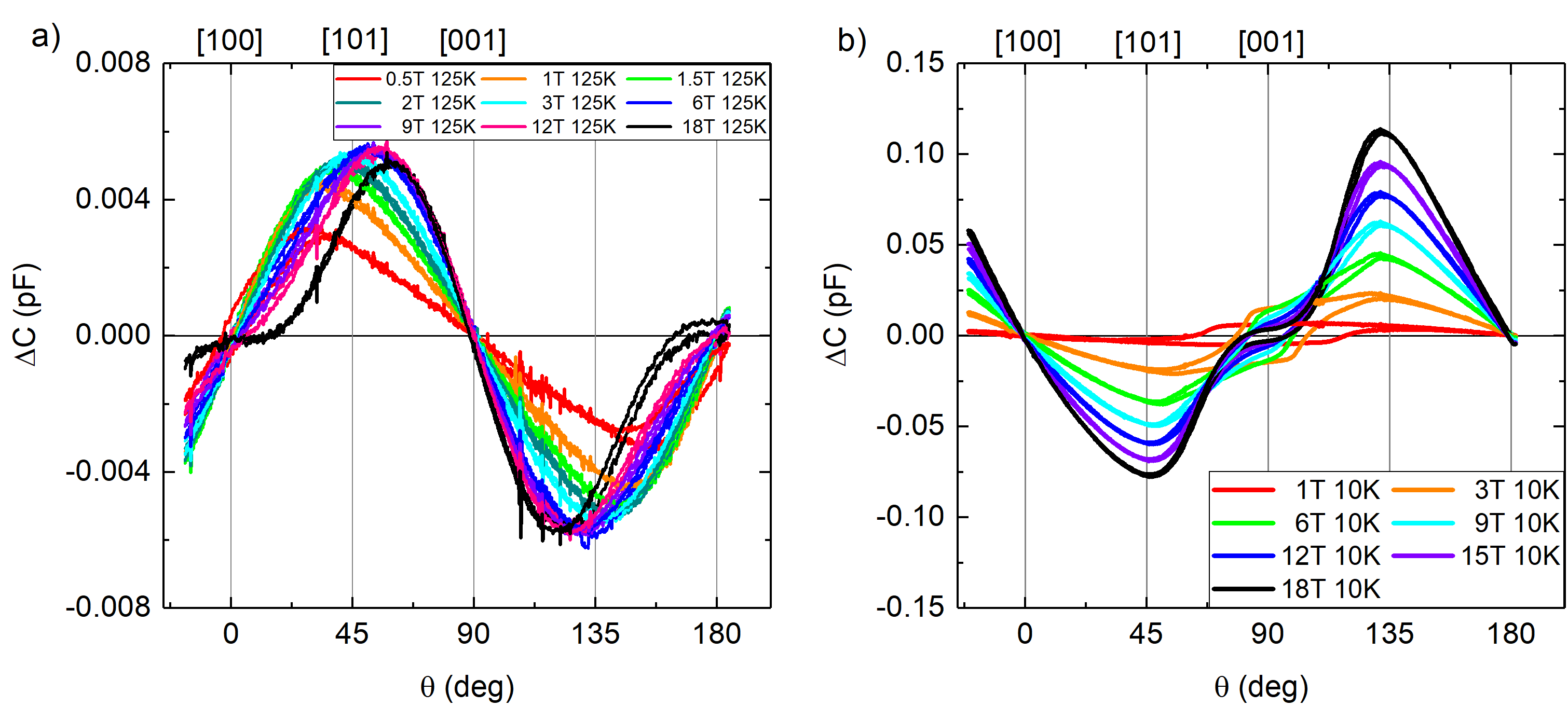}
    \caption{Torque rotations ($\Delta$C vs. $\theta$) for the \hkl(010) rotation plane of a 260 nm thin film, at fields between 0.5 - 18~T. Measurements were performed at a) 125~K (perpendicular anisotropy phase) and b) 10~K (planar anisotropy phase). A $\sin(2\theta)$ response supports uniaxial anisotropy in this rotation plane, and hysteresis around the \hkl[001] direction in the 10~K measurements further supports an out-of-plane hard axis at low temperatures.} 
    \label{OOPRotationsOverviewSM}
\end{figure}

\begin{figure}
    \centering
    \includegraphics[width = 7.0 in]{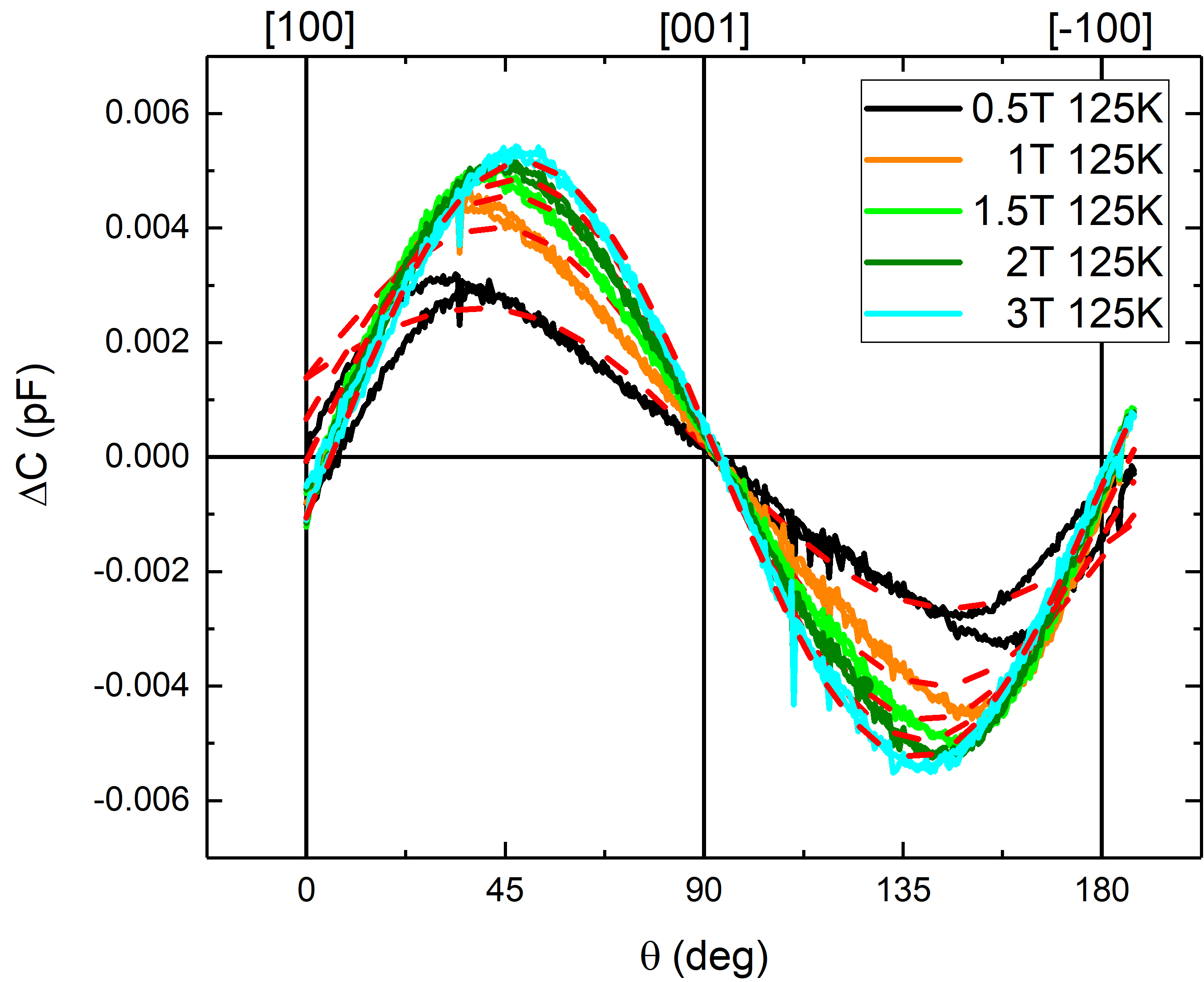}
    \caption{Torque rotations ($\Delta$C vs. $\theta$) for the \hkl(010) rotation plane of a 260 nm thin film, at fields below 3~T taken at T = 125~K. The red dashed lines are fits. }
    \label{LowHfits}
\end{figure}

Fig. \ref{OOPRotationsOverviewSM} shows the \hkl(010) torque curves taken at various fields between $B$~=~0.5~-~18~T. The same curves measured at 3, 6, and 18~T are presented in Fig. 2 of the main text. Panel a) presents the curves taken at 125~K while panel b) shows the curves taken at 10~K. Fig. \ref{LowHfits} shows the low field curves taken at 125~K and their fits using A$_1$sin(2$\theta$) + A$_2$sin(4$\theta$) fit function. The fit results are reported on in Fig. 3 of the main text. 

\begin{figure}
    \centering
    \includegraphics[width = 7 in]{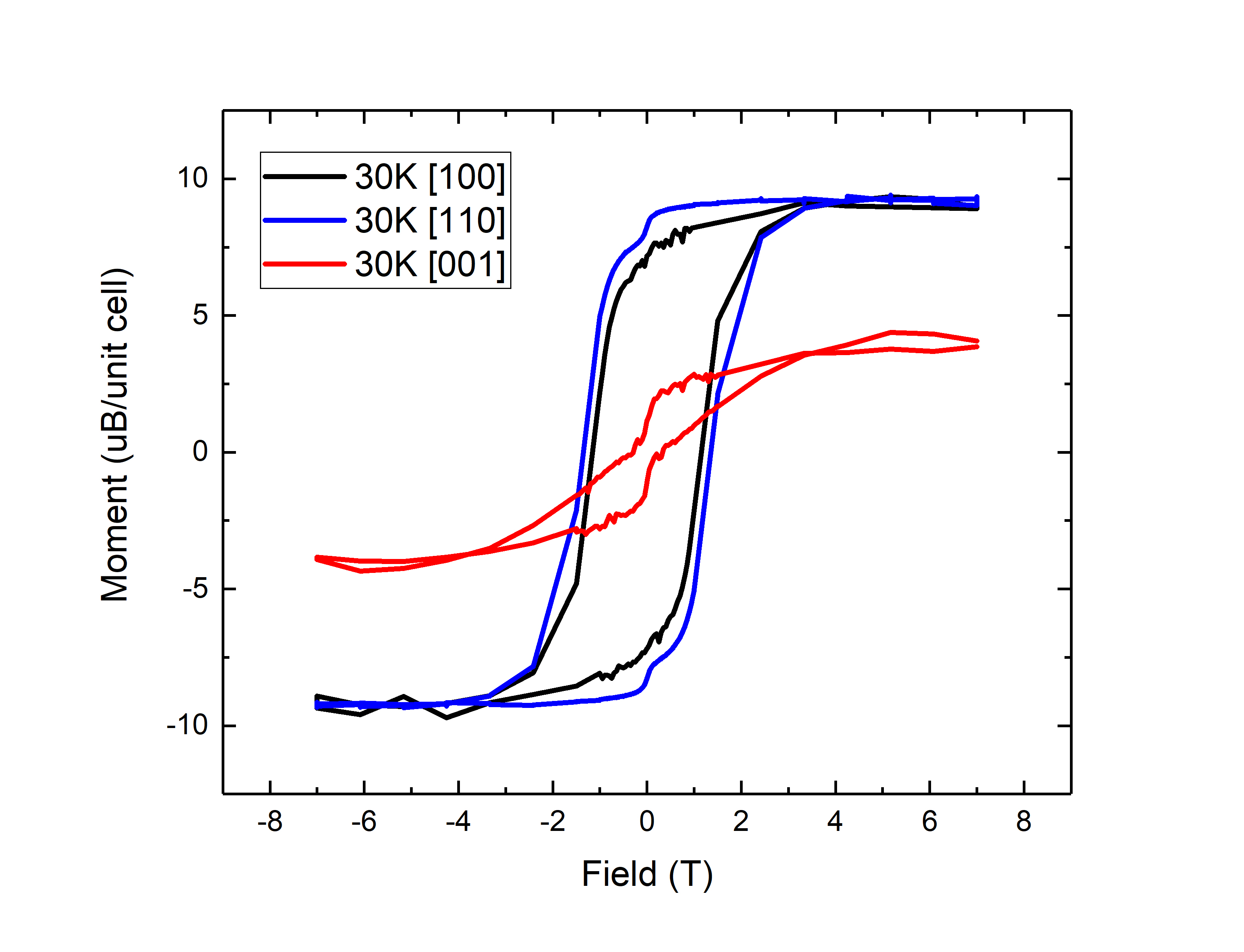}
    \caption{Magnetization as a function of applied field for a 75 nm thin film. Curves are shown for fields directed along the \hkl[100] (black curve), \hkl[110] (blue curve), and \hkl[001] (red curve) direction.  }
    \label{MvsH}
\end{figure}

Magnetization measurements as a function of applied field are shown in Fig. \ref{MvsH} for a 75 nm thick film. Curves are shown with the applied field directed along the \hkl[100], \hkl[110], and \hkl[001] directions. The magnetization saturates most quickly along the \hkl[110] direction, which is in line with our previously reported neutron experiments \cite{Christie}. The double step feature around zero field, visible for the \hkl[110] direction will be further explored in the future, and falls outside the scope of this manuscript. From these measurements we extract the saturated magnetic moments of 9.167 $\pm$ 0.238 $\mu_B$ per unit cell ($B||$\hkl[100]), 9.191 $\pm$ 0.111 $\mu_B$ per unit cell ($B||$\hkl[110]), and 3.904 $\pm$ 0.267 $\mu_B$ per unit cell ($B||$\hkl[001]). 
The magnetization in the \hkl[001] direction shows a much lower value, indicating that saturation is not reached. Saturation magnetization values extracted for other samples scale with the sample thickness, and show similar values compared to the ones reported in Fig. \ref{MvsH}. 



\begin{figure}
    \centering
    \includegraphics[width = 7 in]{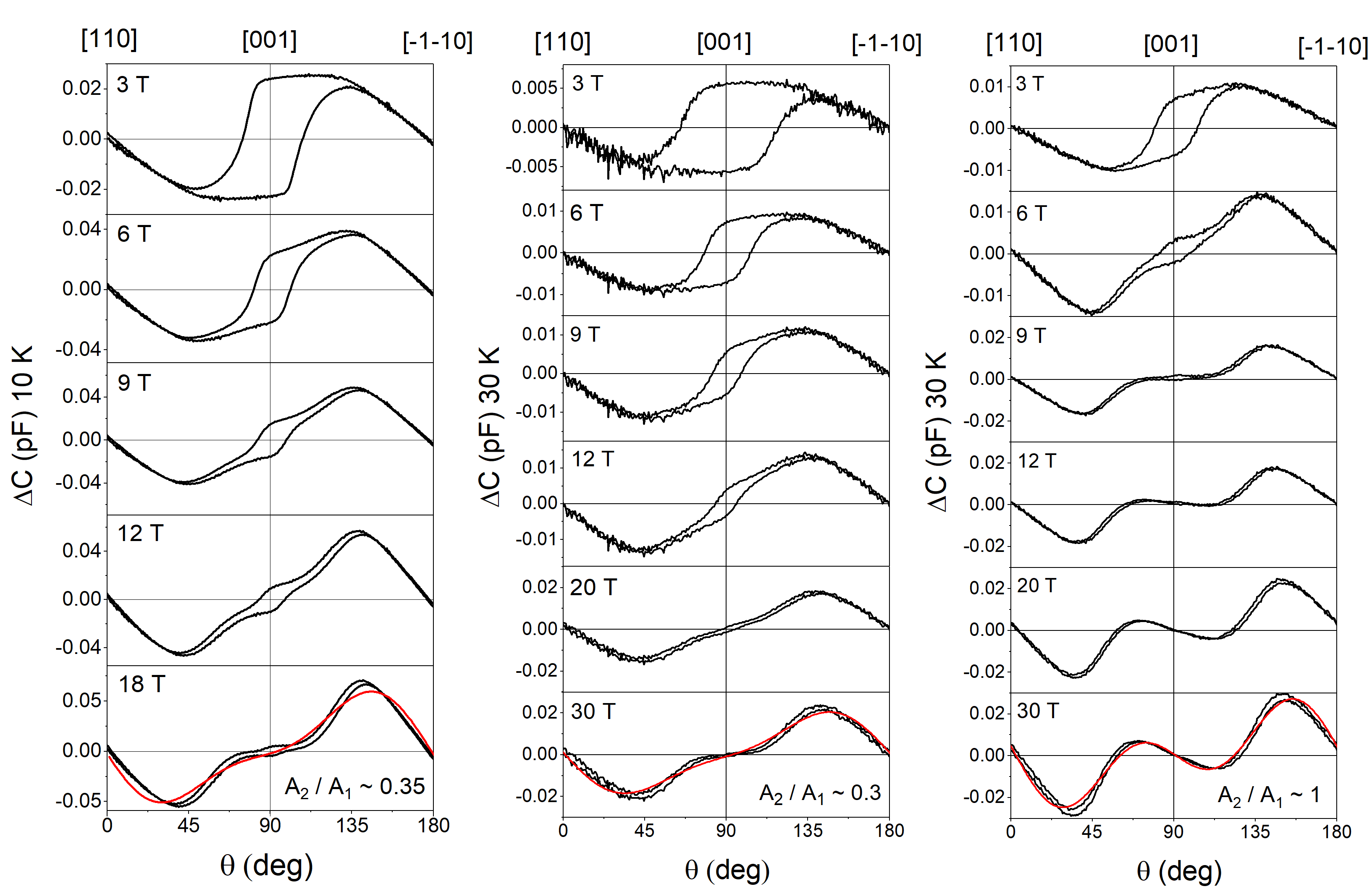}
    \caption{Torque rotations ($\Delta$C vs. $\theta$) for the \hkl(-110) rotation plane (containing \hkl[110] direction and the out of plane \hkl[001] direction) for three different samples taken at fields ranging between 3 - 30~T. Measurements were performed at 10~K (left panel,  for a 260 nm sample) or 30~K (center panel, for a 75 nm sample, and right panel, for a 140 nm sample). In the bottom panel, fit curves (red lines) using A$_1$sin(2$\theta$) + A$_2$sin(4$\theta$) are plotted alongside the data. The A$_2$/A$_1$ ratios indicate the strength of the biaxial term relative to the uniaxial contribution to the magneto-crystalline anisotropy as extracted from the fits. 
    }
    \label{rot-overview}
\end{figure}

Fig. \ref{rot-overview} shows torque rotations ($\Delta$C vs. $\theta$) for the \hkl(-110) rotation plane, which contains the planar \hkl[110] and the out of plane \hkl[001] directions. It is shown for three different samples taken at fields ranging between 3 - 30~T. Measurements were performed at 10 or 30~K. In the bottom panel, fit curves (red lines) using a uniaxial-biaxial model A$_1$sin(2$\theta$)+A$_2$sin(4$\theta$) are plotted alongside the data. The A$_2$/A$_1$ ratios indicate the strength of the biaxial term relative to the uniaxial contribution to the magneto-crystalline anisotropy. For the sample presented in the left panel, which is the same sample for which we presented the \hkl(010) rotations in the Fig. 2 of the main text, this rotation plane shows a similar development of the biaxial term (A$_2$/A$_1$ $\sim$0.35) compared to the measurements taken at 125~K. So, while the overall anisotropy is dramatically different between the collinear and the non-collinear magnetic states, the amount of field-induced magnetostriction with increasing magnetic field applied along the respective hard axes appears to be similar for these different states. In the center and right panel \hkl(-110) torque curves are presented for a 75 and 140~nm sample, respectively, in applied fields up to 30~T. Comparing the bottom panels of Fig. \ref{rot-overview}, we conclude that the variations in high-field biaxial contributions appear to be caused by sample-to-sample changes in magnetostriction effects, which could be due to subtle changes in stoichiometry or variations in sample thickness. 

\begin{figure}
    \centering
    \includegraphics[width = 7 in]{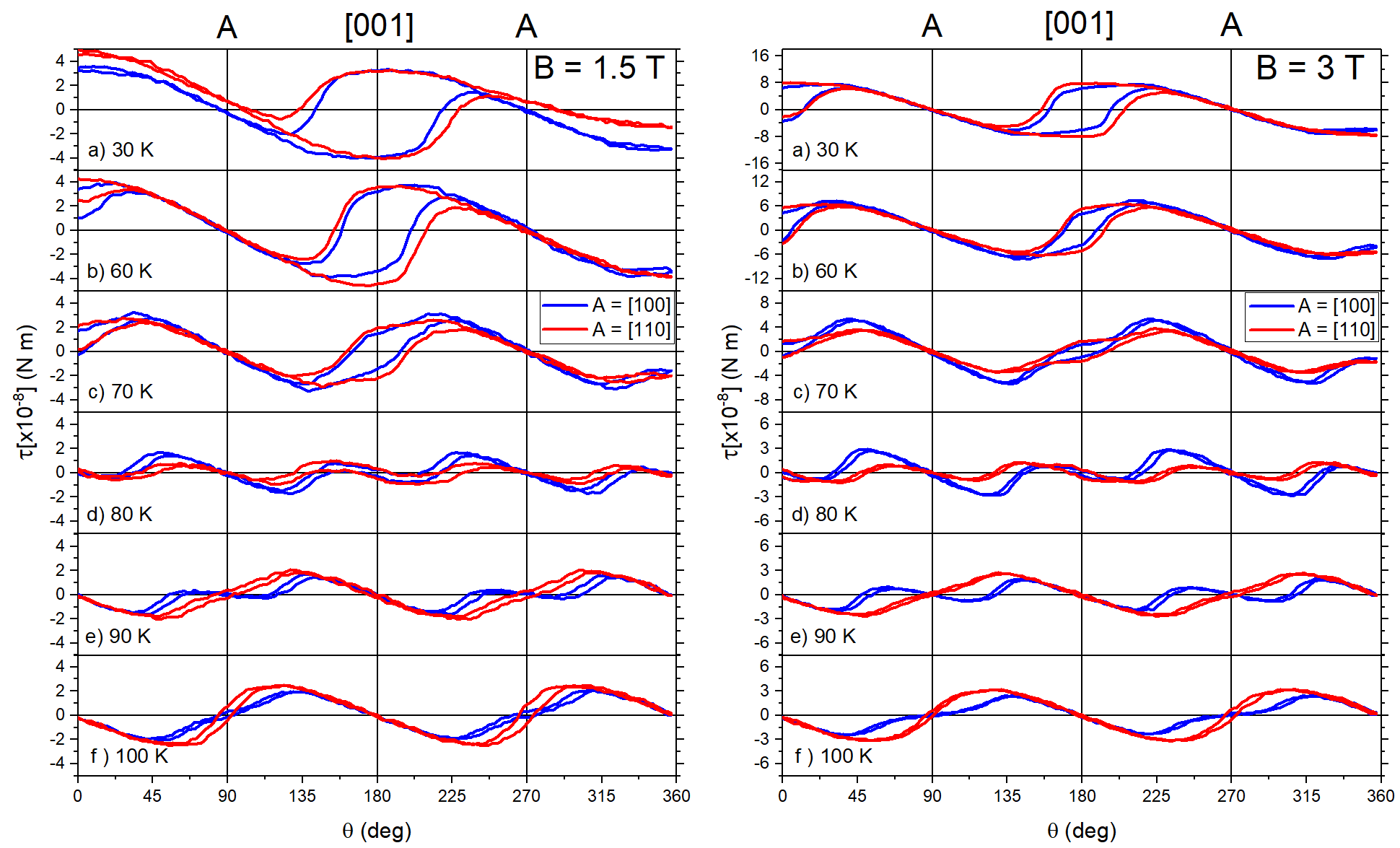}
    \caption{Torque rotations for two rotation planes (shown in red: \hkl(-110) and shown in blue: \hkl(010)) measured at 1.5 and 3~T at temperatures of 30~K, 60~K, 70~K, 80~K, 90~K, and 100~K. Vertical gray line labeled [001] represents when the magnetic field points out-of-plane. The letter ``A" represents when the magnetic field points along the planar \hkl<110> and \hkl<100> directions for the red and blue curves, respectively.  }
    \label{Tempevo}
\end{figure}

In Fig. \ref{Tempevo} we compare the temperature evolution of the torque response at 1.5 and 3~T to highlight the changes in relative anisotropy between the \hkl[001] and the planar \hkl[100] and \hkl[110] directions as the field is lowered. The in-plane directions for each of the rotations is marked with the letter ``A". The temperature evolution, a gradual change from uniaxial to a biaxial response, and then to a uniaxial response, but with an out-of-plane easy axis, is the same as presented in the main text and does not appear to be field dependent. However, from these curves it becomes clear that with decreasing field the hysteresis around the low temperature \hkl[001] hard axis becomes more severe, as expected, and the curves for the \hkl(010) and \hkl(-110) planes become more equivalent at low field, in that their respective amplitudes are the same. This indicates that as the field is lowered the low temperature difference in relative anisotropy between the \hkl[100] and \hkl[110] directions diminishes. Extrapolating, this shows that the torque measurements are not in disagreement with the zero-field neutron measurements, rather the magneto-crystalline anisotropy is very sensitive to applied field due to substantial magnetostriction even at very low fields.

\begin{figure*}
    \centering
    \includegraphics[width = 7 in]{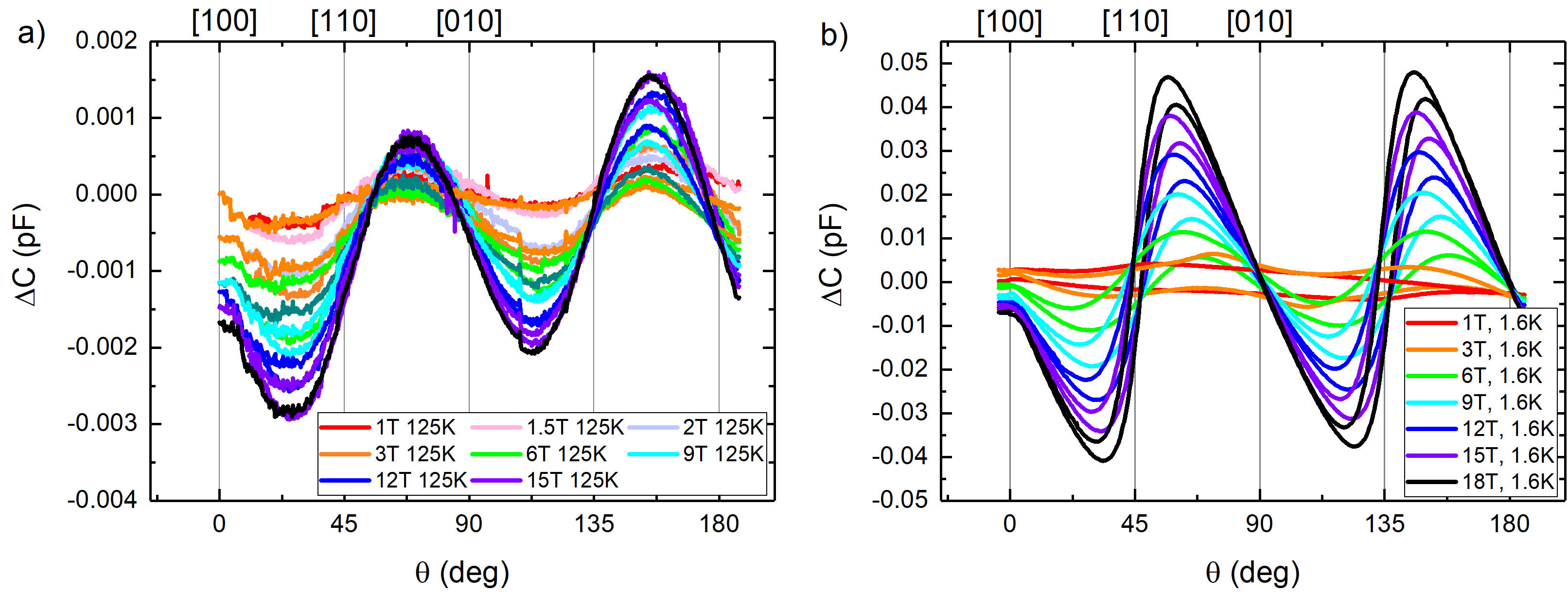}
    \caption{Torque rotations ($\Delta$C vs. $\theta$) for the \hkl(001) plane, at fields between 1 - 18~T. Measurements were performed at a) 125~K (perpendicular anisotropy phase) and b)1.6~K (planar anisotropy phase). A $\sin(4\theta)$ response supports biaxial anisotropy in this rotation plane. 
    }
    \label{IP}
\end{figure*}

\begin{figure}
    \centering
    \includegraphics[width = 7.0 in]{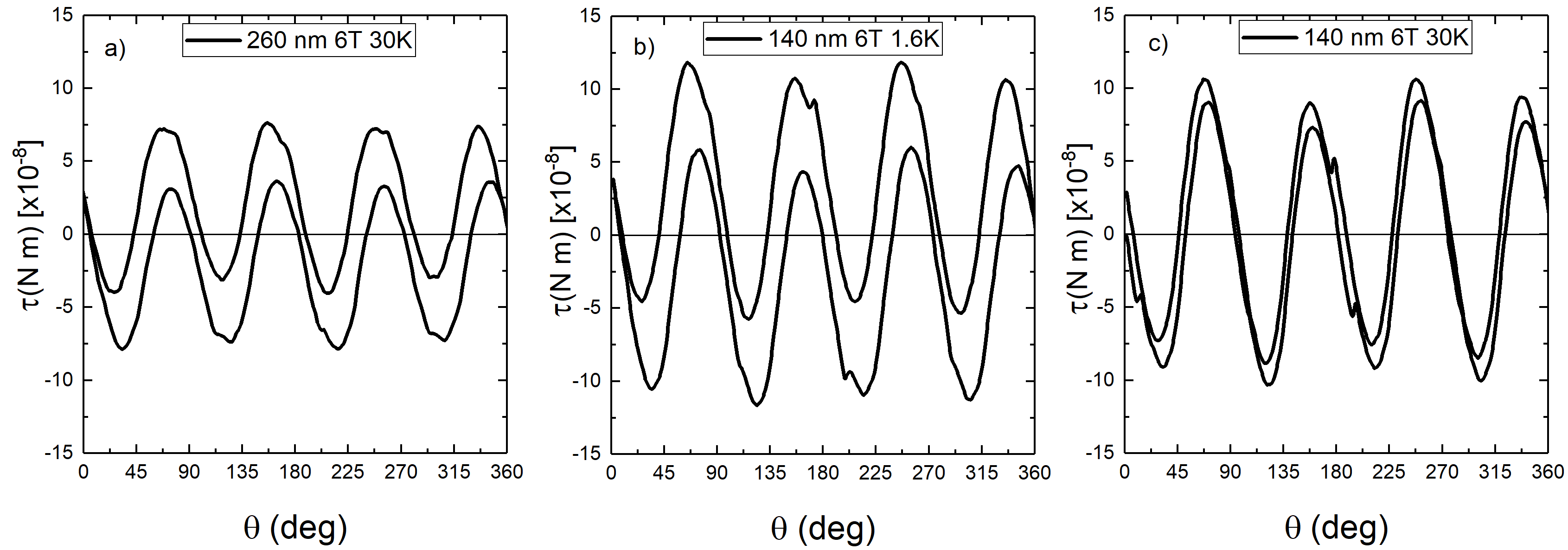}
    \caption{Torque rotations in the \hkl(001) plane for the 260 nm (taken at $T$ = 30~K) and the 140 nm thin film (taken at $T$ = 1.6 and 30~K). These are the same films as shown in the left and right panels of Fig. \ref{rot-overview}.} 
    \label{IPcomp}
\end{figure}

For the 140 nm thin film samples we presented the low temperature \hkl(001) rotations in the main text Fig. 6 and in Fig. \ref{IP} of this Supplemental Materials. Fig. \ref{IP} shows an overview of the torque response for the \hkl(001) rotation plane taken at 125 and 1.6~K. In panel a) the curves taken at 125~K show a biaxial response. The apparent slope is due to a small misalignment of the field introducing a small sin($\theta$) contribution to the signal due to a torque from the \hkl[001] easy axis. Panel b) shows the curves taken at 1.6~K, also showing a biaxial response with hysteresis that is most severe around the \hkl[110] direction, indicating this is a rather hard axis in applied fields. Note, this axis is the zero-field easy axis. In Fig. \ref{IPcomp} we compare the torque response for two different samples (a 140 and a 260 nm film, same samples as in Fig. \ref{rot-overview} left and right panels). The thinner film shows the same in-plane response compared to the thicker sample, but at lower temperatures. This indicates the smaller overall anisotropy in this thinner sample, which is consistent with the sample-to-sample variations in anisotropy and magnetostriction effects mentioned above and in the main text.

\begin{figure}
    \centering
    \includegraphics[width = 6 in]{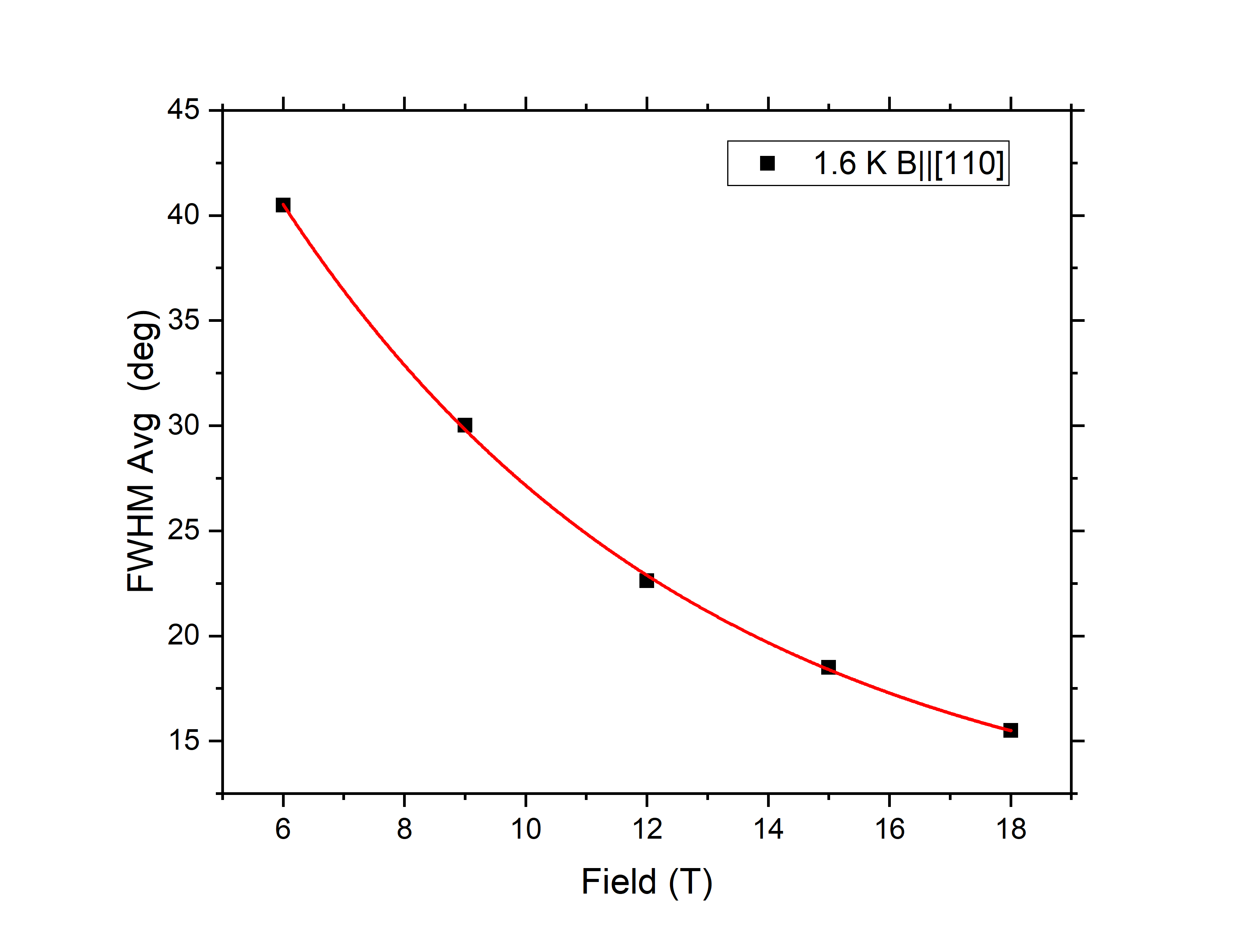}
    \caption{Exponential decay fits of FWHM values of Zeeman angle plots for the in-plane rotation, with fields along the \hkl[110] direction. Temperature set at 1.6~K. Extrapolating the fitting parameters estimates a Zeeman angle of 1$\degree$ at 30~T.}
    \label{IPExp}
\end{figure}

Fig. \ref{IPExp} shows the exponential decay of the Full-Width Half-Maximum (FWHM) values of the in-plane rotation Zeeman angle. Fits were done to a standard Gaussian distribution, with FWHM values from both peaks at 45$\degree$ and 135$\degree$ in Fig. 6c) in the main text averaged out to give the final value for the exponential decay. The fit of the exponential curve returns a Zeeman angle of $\sim$1$\degree$ at 30~T.

\begin{figure}
    \centering
    \includegraphics[width = 7 in]{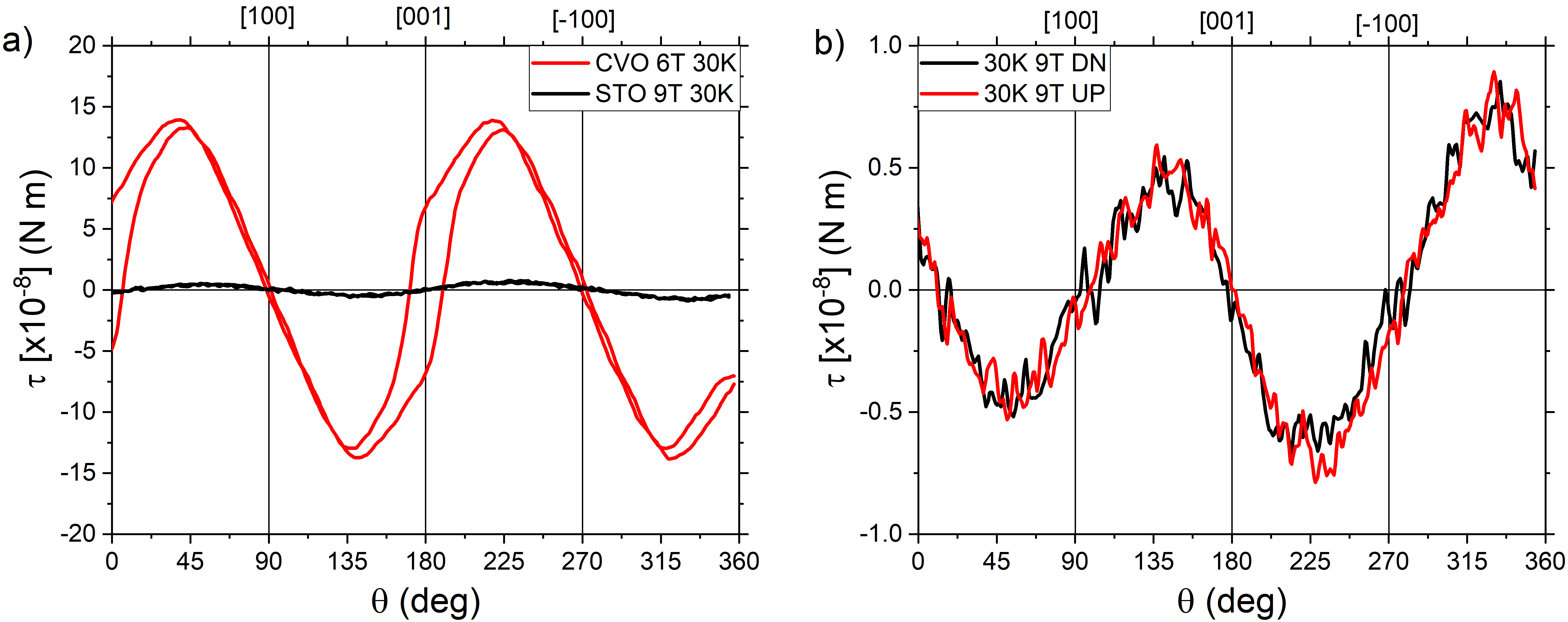}
    \caption{a) Torque magnetometry curves of CoV$_2$O$_4$ thin film and the SrTiO$_3$ substrate for the \hkl(010) rotation plane. Both measurements were done at 30~K, with the CoV$_2$O$_4$ thin film measurement performed at 6~T and the SrTiO$_3$ measurement was done at 9~T. b) The same SrTiO$_3$ response as plotted in the left panel, but zoomed in. No hysteresis is seen for the SrTiO$_3$ when overlaying forward and backward rotations (black and red curves).}
    \label{STOback}
\end{figure}

Fig. \ref{STOback} demonstrates the substrate contribution to the torque signal at $T$~=~30~K and in $B$~=~9~T. The substrate response is plotted alongside the CoV$_2$O$_4$ response measured at the same temperature but in a lower field of 6~T (left panel). The right panel shows a zoomed in plot of the same substrate torque response.  Note the measurement accuracy of the PPMS Torque magnetometry option is limited to 10$^{-9}$ N m. Thus, the substrate response is close to the sensitivity limit and almost negligible compared to the response from the film.

\bibliography{references.bib}